\documentclass[aps,nofootinbib,amsmath,prd,twocolumn,showpacs,superscriptaddress,groupedaddress]{revtex4-1}

\usepackage{graphicx}  % needed for figures
\usepackage{dcolumn}   % needed for some tables
\usepackage{bm}        % for math
\usepackage{amssymb}   % for math

\usepackage[utf8]{inputenc}
\usepackage{slashed}
\usepackage{epsfig}
\usepackage{amsmath,amsfonts,mathtools}  
\usepackage{color}

\begin{document}

\mbox{}

\title{Scaling and superscaling solutions from the functional renormalization group}

\author{Tobias Hellwig}
\email{tobias.hellwig@uni-jena.de}
\affiliation{Theoretisch-Physikalisches Institut, Friedrich-Schiller-Universit\"{a}t Jena,\\
Max-Wien-Platz 1, 07743 Jena, Germany}

\author{Andreas Wipf}
\email{wipf@tpi.uni-jena.de}
\affiliation{Theoretisch-Physikalisches Institut, Friedrich-Schiller-Universit\"{a}t Jena,\\
Max-Wien-Platz 1, 07743 Jena, Germany}

\author{Omar Zanusso}
\email{omar.zanusso@uni-jena.de}
\affiliation{Theoretisch-Physikalisches Institut, Friedrich-Schiller-Universit\"{a}t Jena,\\
Max-Wien-Platz 1, 07743 Jena, Germany}

% \date{\today}

\begin{abstract}
We study the renormalization group flow of $\mathbb{Z}_2$-invariant 
supersymmetric and non-supersymmetric scalar models
in the local potential approximation using functional renormalization group methods.
We focus our attention on the fixed points of the renormalization group flow of these models,
which emerge as scaling solutions.
In two dimensions these solutions are interpreted as the minimal (supersymmetric) models of conformal field theory,
while in three dimensions they are manifestations of the Wilson-Fisher universality class and its supersymmetric counterpart.
We also study the analytically continued flow in fractal dimensions between $2$ and $4$ and determine the critical dimensions
for which irrelevant operators become relevant and change the universality
class of the scaling solution.
We include novel analytic and numerical investigations of the properties that determine the occurrence of the scaling solutions within the method.
For each solution we offer new techniques to compute the spectrum of the deformations
and obtain the corresponding critical exponents.
\end{abstract}

\maketitle

\section{Introduction}
\label{section_intro}

The birth of functional renormalization group methods is tightly connected
to the investigation of scaling properties of the effective potential in 
statistical field theories with $\mathbb{Z}_2$ or ${\rm O}(N)$ symmetries \cite{ZinnJustin:2002ru,Pelissetto:2000ek},
and it represents its first non-trivial application \cite{Wetterich:1992yh,Morris:1993qb,Berges:2000ew}.
Over the years the interest in properties that can be 
extracted from the renormalization group flow of the potential
in a local approximation has been revitalized
at various stages \cite{Morris:1998da,Rosten:2010vm}.

The effective potential of a $\mathbb{Z}_2$-symmetric scalar field 
theory has been studied with
various techniques to unveil the underlying universality classes, which include the Ising model.
The critical exponents have been determined to high accuracy within polynomial 
expansions \cite{Litim:2001dt,Litim:2002cf,Litim:2003kf,Litim:2010tt} and non-polynomial solutions have been studied 
numerically by applying scaling solutions methods \cite{Morris:2005ck,Bervillier:2007tc}.
Recently, even global solutions have been constructed with very high 
numerical precision \cite{Borchardt:2015rxa}.

In the past few years, the scaling solutions approach was adopted to analytically continue various parameters which enter the
renormalization group flow, including the dimensionality of space \cite{Codello:2012sc}, the number $N$ of fields \cite{Comellas:1997tf,Codello:2012ec,Boettcher:2015pja,Percacci:2014tfa}
or both \cite{Codello:2014yfa}. The analytic continuation of these parameters offers new insights into properties
of the underlying statistical models, such as their emergence on spaces of fractional dimensions, the occurrence of the Mermin-Wagner theorem
and constraints on the existence of the various multicritical models.

In this paper we extend the scaling solution approach in three prominent ways.
First, we will carefully lay down and discuss the theoretical mechanism with which a scaling solution occurs
for a specific statistical model, which has often been overlooked in previous works.
This occurrence will be illustrated mostly through the example of the three dimensional Ising model.
Second, we will introduce two methods to compute the critical 
exponents of a model:
a shooting approach (which has been used previously in \cite{Morris:1994ie,Morris:1996kn,Morris:1994ki}) to calculate critical 
exponents in three and two dimensions)
and the novel pseudo-spectral method with a SLAC derivative
\cite{Drell:1976bq,Kirchberg:2004vm,Wipf:2013vp}. 
These two methods treat the boundary conditions
for the deformations of a critical potential in different ways; thus, they are complementary for the approach
as will be shown by applying them to the known Ising example.
Finally, we systematically apply the scaling solution approach
to the supersymmetric ${\cal N}=1$ Wess-Zumino model in both two \cite{Synatschke:2009nm,Mastaler:2012xg} and three dimensions \cite{Synatschke:2009da,Synatschke:2010ub,Heilmann:2012yf},
for which we compute various critical exponents for its universality classes.
We also study the supersymmetric model on fractional dimensions
introducing a novel analytic continuation. % for it.

The paper is organized as follows:
The remaining part of Section \ref{section_intro} introduces the local potential approximation
and the functional renormalization approach,
Sect.\ \ref{section_simple_scalar} introduces the scaling solution approach
and applies it to various universality classes in two and three dimensions,
Sect.\ \ref{section_wz_model} introduces the Wess-Zumino model
and shows the application of the scaling solution method to the supersymmetric analogues
of the universality classes of the previous Section,
Sect.\ \ref{section_critical_dimensionalities}
shows how the universality classes emerge at certain critical dimensions
when varying the dimensionality of the system,
 and finally Sect.\ \ref{section_discussion} briefly discusses our results.
The appendixes are dedicated to further clarifying some properties of the renormalization group flow
and the methods used in this work.

% \vspace{-0.209cm}

\subsection{Local potential approximation}
% \label{intro}

For clarity purposes we introduce all the basic concepts used in the present paper
adopting the illustrative example of a simple $\mathbb{Z}_2$-invariant scalar field on $\mathbb{R}^d$.
This introduction can and will be straightforwardly generalized to the inclusion of supersymmetry later
in the corresponding Sections.
Readers that are experts in the techniques of Wilsonian and functional renormalization are welcome to skip the remainder of the present section.

The local potential approximation (LPA) is a non-perturbative truncation of the effective action of a system
which includes all the relevant operators in a low energy derivative expansion \cite{Nicoll:1974zz,Bervillier:2013hha}.
Within this approximation, higher derivative operators are neglected but all local interactions are included in a non-perturbative fashion.
The LPA of a simple $\mathbb{Z}_2$-invariant scalar $\varphi$ in $d$ dimensions thus truncates the effective action of the system to
\begin{equation}\label{scalar_effective_action}
 \Gamma[\varphi] = \int {\rm d}^d x \left\{ \frac{Z}{2}\partial_\mu \varphi \partial^\mu\varphi + V(\varphi) \right\}\,,
\end{equation}
where we included a wavefunction renormalization constant $Z$ for the field $\varphi$ and the local potential $V(\varphi)$.
More precisely, the truncation \eqref{scalar_effective_action} represents an extension of the original LPA of a simple scalar
that goes under the name of improved LPA (a.k.a.\ LPA'),
as it includes a non-trivial wavefunction renormalization for $\varphi$ and thus allows the field to have a non-canonical scaling
in the guise of an anomalous dimension for the renormalized field $\varphi_{\rm R}\equiv Z^{1/2} \varphi$.

\subsection{Functional Renormalization}

We study the renormalization group flow of the scalar models using functional renormalization group (FRG) methods.
In this approach the path-integral of the system is constructed as
\begin{equation}\label{eaa_path_integral}
 {\rm e}^{-\Gamma_k[\varphi]} = \int {\cal D}\chi \, {\rm e}^{-S[\chi]-\Delta S_k[\chi] +(\chi-\varphi)\frac{\delta\Gamma_k}{\delta\varphi}}\,,
\end{equation}
where $S[\chi]$ is a bare action for the microscopic field $\chi$
and $\Delta S_k[\chi]$ is a cutoff that suppresses the propagation of the infrared modes 
relative to a given reference scale $k$.
The infrared cutoff should be quadratic in the field
\begin{equation}\label{cutoff}
 \Delta S_k[\chi] = \frac{1}{2} \int {\rm d}^d x \, {\rm d}^d x' \, \chi(x) {\cal R}_k(x-x') \chi(x')\,,
\end{equation}
and its kernel ${\cal R}_k$ suppresses the propagation of the infrared modes by directly modifying the bare propagator of the theory.
The dependence of \eqref{eaa_path_integral} on the kernel ${\cal R}_k$ encodes the scheme-dependence of the method;
thus, different kernels implement different coarse-grainings.
However, the kernel has to satisfy in the limit $k\to 0$ the property ${\cal R}_{k\to 0}=0$,
which ensures that
\eqref{eaa_path_integral} reduces to the standard path-integral of the (full) effective action $\Gamma[\varphi]$ of the system
 when the reference scale $k$ is taken to zero.

The functional $\Gamma_k[\varphi]$ appearing on the rhs of \eqref{eaa_path_integral} is known as the effective average action (EAA) of the model.
By construction it reduces to the full effective action $\Gamma_{k\to 0}[\varphi]=\Gamma[\varphi]$ in the infrared limit $k\to 0$.
Furthermore, for sufficiently regular cutoff kernels it can be 
shown that an ultraviolet limit can be constructed
$\Gamma_{k\to \Lambda}[\varphi]=S_\Lambda[\varphi]$, in which $\Lambda$ is an 
ultraviolet cutoff
and $S_\Lambda[\varphi]$ is the (possibly divergent) bare action of the model 
at the UV-scale $\Lambda$.
While the path integral \eqref{eaa_path_integral} is generally not free of ultraviolet divergences and, in principle, requires renormalization,
it is possible to show that the flow of $\Gamma_k[\varphi]$ with respect to the scale $k$ is finite.
Therefore, rather than achieving the renormalization of the theory through the inclusion of counterterms,
it is customary to use \eqref{eaa_path_integral} to derive the exact 
and finite FRG equation
\begin{equation} \label{erge}
 k\partial_k \Gamma_k[\varphi]
 =
 \frac{1}{2} {\rm Tr} \left(\Gamma^{(2)}_k[\varphi]+{\cal R}_k\right)^{-1}k\partial_k {\cal R}_k\,,
\end{equation}
and use it to investigate the renormalization of the model.

The flow encoded by \eqref{erge} determines a velocity field in the space 
of all functionals of the fields entering a model under investigation.
The LPA can be applied easily by projecting both sides of \eqref{erge} to a $k$-dependent generalization of \eqref{scalar_effective_action} which reads
\begin{equation}\label{scalar_effective_average_action}
 \Gamma_k[\varphi] = \int {\rm d}^d x \left\{ \frac{Z_k}{2}\partial_\mu \varphi \partial^\mu\varphi + V_k(\varphi) \right\}\,.
\end{equation}
The anomalous dimension of the field is $\eta \equiv -k\partial_k Z_k/Z_k$.
In order to read off the renormalization group flow of the couplings in \eqref{scalar_effective_average_action}
we choose the prescriptions
\begin{equation}\label{extraction_scheme}
 \begin{split}
  & k\partial_k V_k(\varphi)
  =
  \frac{1}{\rm Vol} \, k\partial_k \!\left. \Gamma_k[\varphi] \right|_{\varphi={\rm const.}}\,,
  \\
  &\eta = -\frac{1}{Z_k} k\partial_k  \frac{\partial}{\partial p^2} \left.\frac{\delta^2 \Gamma_k[\varphi]}{\delta \varphi_p \delta\varphi_{-p}}\right|_{\varphi=\varphi_0,\,p^2=0}
 \end{split}
\end{equation}
which must be applied to the right hand side of \eqref{erge}.
Various remarks are in order to explain the meaning of \eqref{extraction_scheme}:
The flow of the local potential is defined as the (normalized) flow of the EAA in the limit of constant field content,
as is usual in quantum field theoretical applications.
The field $\varphi_p$ represents the Fourier transform of $\varphi(x)$ with respect to the momentum $p$;
thus, the anomalous dimension $\eta$ is extracted quite naturally from the coefficient of the $p^2$ term in the two-point function belonging to $\Gamma_k[\varphi]$.
Since the truncation \eqref{scalar_effective_average_action} includes only a field-independent wavefunction renormalization constant,
it is necessary for consistency to project the two-point function 
of $\Gamma_k[\varphi]$ to a distinguished field configuration $\varphi_0$.
In applications $\varphi_0$ is chosen in a physically meaningful way as 
the minimum of the effective potential defined by $V_k'(\varphi_0)=0$.
The configuration $\varphi_0$ thus plays the role of the order parameter of the $\mathbb{Z}_2$-symmetry $\varphi_0=\left<\varphi\right>$,
and the possibility that it has non-zero values represents a clear separation from standard perturbation theory.

Let us now define the modified momentum-space propagator at constant field
\begin{equation}
 {\cal G}_k(q^2) \equiv
 \left(Z_k q^2+ V_k''(\varphi)+ {\cal R}_k(q^2)\right)^{-1}\,,
\end{equation}
where ${\cal R}_k(q^2)$ is $(2\pi)^d$ times the Fourier transform of the cutoff kernel in coordinate space.
With the modified propagator we can give a compact evaluation of \eqref{extraction_scheme} based on a general cutoff
\begin{equation}\label{general_simple_scalar_flow}
 \begin{split}
  &k\partial_k V_k(\varphi)
  =
  \frac{1}{2(2\pi)^d} \int {\rm d}^dq \, {\cal G}_k \, k\partial_k {\cal R}_k
  \\
  &\eta = -\frac{V^{(3)}(\varphi_0)^2}{(2\pi)^d Z_k} \int {\rm d}^dq \left.\left({\cal G}_k'+q^2\frac{2}{d}{\cal G}_k''\right){\cal G}_k^2\, k\partial_k {\cal R}_k\right|_{\varphi_0}
 \end{split}
\end{equation}
Here, the prime denotes the derivative with respect to the argument $q^2$,
whereas elsewhere it is the derivative with respect to the (constant)
field.

\section{Scalar field theory}

\label{section_simple_scalar}

We can give explicit expressions for the flow equation for the effective potential and the anomalous dimension after specifying the cutoff function.
While, in principle, any choice compatible with the properties of a suitable cutoff is possible,
for the purpose of this paper it is useful to choose one that allows for
an explicit calculation of the flow.
We choose the popular optimized cutoff and explicitly factor out the wavefunction renormalization of the field, so that in momentum space the kernel of \eqref{cutoff} reads
\begin{equation}\label{cutoff_optimized}
 {\cal R}_k(q^2) = Z_k (k^2-q^2)\theta(k^2-q^2)\,.
\end{equation}
The optimized cutoff is a distribution and therefore the momentum space integrals \eqref{general_simple_scalar_flow}
must be treated with care.\footnote{The presence of the derivatives with respect to $q^2$ in \eqref{general_simple_scalar_flow}
requires a normalization for the integral of the product of the Heaviside step function $\theta(y-y_0)$ and its first derivative $\delta(y-y_0)$.
Given a regular test function $g(y)$ we normalize $$\int {\rm d}y \,g(y)\,\delta(y-y_0)\,\theta(y-y_0)=1/2 \,g(y_0)\,.$$
This normalization is consistent with the limit of the same integral evaluated for any sequence of smooth functions approaching $\theta(y-y_0)$.
}

Let us also introduce the dimensionless renormalized field $\overline{\varphi}_{\rm R}\equiv Z_k^{-1/2} k^{(2-d)/2}\varphi$,
which is the natural argument of the dimensionless renormalized potential
\begin{equation}
 v_k(\overline{\varphi}_{\rm R}) \equiv k^{-d} V_k(\varphi)\,.
\end{equation}
Together $\overline{\varphi}_{\rm R}$ and $v_k(\overline{\varphi}_{\rm R})$ are the appropriate variables to investigate the scaling properties of the system,
and in fact we shall use them for the remainder of this section.
For notational simplicity we also drop both the labels and the overline notation, so the explicit evaluation of the flow in the new variables takes the form
\begin{equation}
 \begin{split}\label{vdot}
  &k\partial_k v(\varphi)
  =
  {\cal S}_{v,\eta}[v,v';\varphi]+{\cal F}_{v,\eta}[v'']\,,
  \\
  &{\cal S}_{v,\eta}[v,v';\varphi]\equiv-d v(\varphi)+\frac{d-2+\eta}{2}\varphi v'(\varphi)\,,
  \\
  &{\cal F}_{v,\eta}[v'']\equiv c_d \frac{1-\frac{\eta}{d+2}}{1+v''(\varphi)}\,,
 \end{split}
\end{equation}
in which we defined the dimension-dependent geometric constant $c_d^{-1}\equiv (4\pi)^{d/2}\Gamma(1+d/2)$.
We split the flow of the potential in two parts with different origins.
The functional ${\cal S}_{v,\eta}[v,v';\varphi]$ represents the scaling part of the flow
and it originates from the change of dimensionful variables to dimensionless ones.
Its form is independent from the cutoff choice \eqref{cutoff_optimized} and 
therefore from the scheme of computation. 
The functional ${\cal F}_{v,\eta}[v'']$, instead, encodes the nontrivial features of the flow and depends on the cutoff.

We now compute from \eqref{general_simple_scalar_flow} the anomalous dimension explicitly to obtain
\begin{equation}\label{eta}
 \eta = c_d \frac{v'''(\varphi_0)^2}{(1+v''(\varphi_0))^4}\,,
\end{equation}
where again we assume $\varphi_0$ to be the expectation value of the order 
parameter of the system.
For later convenience we also give the result of an alternative 
computation of the anomalous dimension,
which comes as the limit $N \to 1$ of the anomalous dimension of the Goldstone modes of the ${\rm O}(N)$-model
\begin{equation}\label{goldstone_eta}
 \eta' = 2 c_d \frac{v''(\varphi_0)^2}{\varphi_0^2(1+v''(\varphi_0))^2}\,.
\end{equation}
This definition has often and successfully been used in place of $\eta$ 
on the basis of physical arguments.
Thus, we include it in our analysis and refer to it as Goldstone anomalous dimension.
The reason why $\eta$ and $\eta'$ are different is not because the limit $N\to 1$ of the ${\rm O}(N)$-model is discontinuous,
but rather because $\eta$ represents the limiting value of the anomalous dimension of the massive Higgs field in the broken ${\rm O}(N)$ phase.

\subsection{Scaling solutions}

\label{subsection_scaling_solutions}

Let us now discuss some general property of the flow of the potential \eqref{vdot}.
The computation can be manifestly analytically continued to any non-integer dimensionality,
and for the present work we are mostly concerned with the interval $2\leq d \leq 4$
which is bounded by the lower and upper critical dimensions of the $\varphi^4$ model.
Note that the flow of the potential depends \emph{parametrically} on 
the anomalous dimension given in \eqref{eta} or \eqref{goldstone_eta}.
Nevertheless, for $0<\eta<1$ we can extract some general properties of the flow
that are insensitive to the values of $\eta$.
The flow \eqref{vdot} is a PDE representing the evolution of the potential with respect to the scale $k$.
The $k$-independent (stationary) solutions are the critical solutions 
(fixed points) of the flow,
and they correspond to critical models of the corresponding field theory.
This means that critical solutions solve the non-linear ODE
\begin{equation}
 {\cal S}_{v,\eta}[v,v';\varphi]+{\cal F}_{v,\eta}[v'']=0\,,
\end{equation}
which is called a fixed point equation.
The function ${\cal F}_{v,\eta}$ can be 
explicitly inverted to cast the ODE in a standard form
\begin{equation}\label{v_ode}
 v''(\varphi) = - {\cal F}^{\,(-1)}_{v,\eta}[\,{\cal S}_{v,\eta}[v,v';\varphi]\,]\,.
\end{equation}
While the explicit form of this ODE depends on the cutoff, it is easy to prove that for any cutoff
\begin{equation}
 {\cal F}_{v,\eta}[x] \sim 1/x\,,\,\,\,\, {\rm for}\,\,\,\, x \to \infty\,,
\end{equation}
which implies
\begin{equation}\label{singularity0}
 v''(\varphi) \sim -1/{\cal S}_{v,\eta}[v,v';\varphi]\,,\,\,\,\, {\rm when}\,\,\,\, {\cal S}_{v,\eta} \sim 0\,.
\end{equation}
In other words the second derivative of the potential is expected to diverge at points $\tilde{\varphi}$
at which the scaling part becomes zero. These points can be determined implicitly by the equation ${\cal S}_{v,\eta}=0$ as
\begin{equation}\label{singularity}
 \tilde{\varphi}= \frac{2d}{d-2+\eta}\frac{v(\tilde{\varphi})}{v'(\tilde{\varphi})}\,.
\end{equation}
It is possible to show
that any attempt to integrate the 
fixed point equation \eqref{v_ode} starting from initial conditions at $\varphi=0$
will stop (with probability one) at a singularity located at $\tilde{\varphi}$ at
which $v''(\tilde{\varphi})$ diverges, but $v(\tilde{\varphi})$ 
and $v'(\tilde{\varphi})$ stay finite \cite{Morris:1994ki}.
In addition, if $\tilde{\varphi}>0$ then $v(\tilde{\varphi})$ 
and $v'(\tilde{\varphi})$ have the same sign;
similarly, if $\tilde{\varphi}<0$
they
% then $v(\tilde{\varphi})$ and $v'(\tilde{\varphi})$ 
have opposite signs.

Ideally, we are interested in global analytic solutions of \eqref{v_ode}, 
which thus avoid the appearance of singularities such as \eqref{singularity} for finite values of the field $\tilde{\varphi}$,
and we hope that these global solutions can be obtained by numerically integrating the ODE \eqref{v_ode} with fine tuned boundary conditions.
In practice, it turns out that in the process of integrating \eqref{v_ode} singularities are the rule and not the exception. 
To investigate the singular solutions we first choose a suitable
parametrization.
Since \eqref{v_ode} is a second order ODE, any solution can be characterized
by the two initial conditions $v(0)$ and $v'(0)$.
One of the conditions is, however, fixed by the requirement 
of $\mathbb{Z}_2$-invariance to be $v'(0)=0$,
thus ensuring the even parity of the solutions.
It proves convenient to parametrize the second condition using the value $v''(0)$ rather than $v(0)$.
We thus choose, as an initial condition $v''(0)=\sigma$, which implies that
\begin{equation}\label{v_zero}
 v(0) = \frac{1}{c_d d}\left(1-\frac{\eta}{d+2}\right)\frac{1}{1+\sigma}\,,
\end{equation}
which makes explicit the fact that $\sigma=-1$ is a singular point for the ODE.
This comes from the more general property that, for any cutoff, the flow is well defined whenever $v''(\varphi)>-1$.
With the exception of the trivial case $\sigma=0$, it is possible to observe numerically that almost all solutions that can be considered within $2\leq d\leq 4$
diverge at a singular point $\tilde{\varphi}$ defined by \eqref{singularity} whose position depends on $\sigma$.
We therefore denote $\tilde{\varphi}_\sigma$ the location of the singularity as a function of $\sigma$.

The second derivative $v''(\varphi)$ diverges at field values for which the scaling
part $S_{v,\eta}$ vanishes. At any maximum or minimum of the potential with 
positive $v$ the scaling part is negative and at any extremum
with negative $v$ the scaling part is positive.
Since for $\sigma>-1$ the scaling part is negative at $\varphi=0$
and since the potential (of a stable theory) is positive for large
values of the field we conclude that \emph{every globally defined
fixed point solution $v(\varphi)$ is positive}. 
If $v$ takes negative values
for certain fields, then it would have a minimum $\varphi_\mathrm{min}$
with negative $v$ and the scaling part would change sign between $\varphi=0$ and 
$\varphi_\mathrm{min}$. This in turn implies that the scaling part 
would vanish at some  field $\tilde{\varphi}_\sigma<\varphi_\mathrm{min}$ or 
that the potential would be singular at $\tilde{\varphi}_\sigma$.

\begin{figure}[tph]
 \includegraphics[width=8cm]{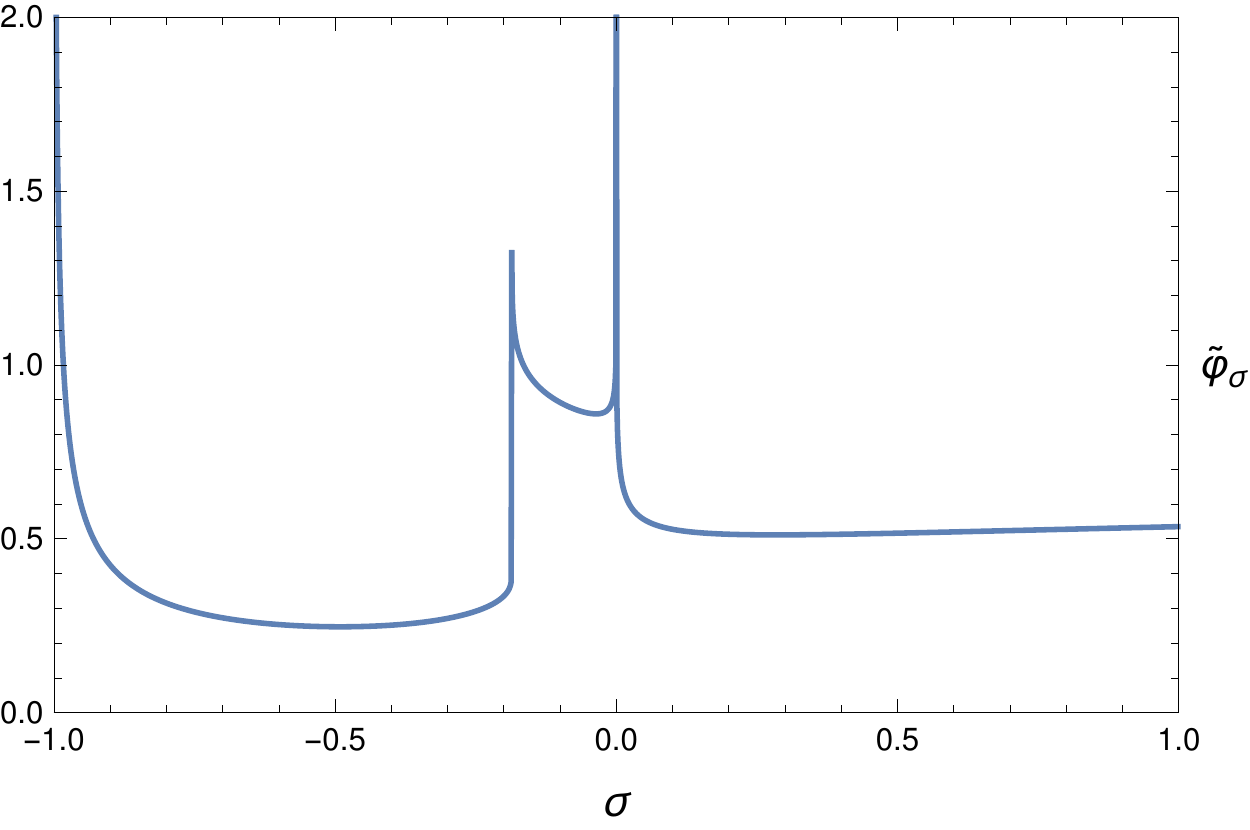}
 \caption{
 Location of the singularity $\tilde{\varphi}_\sigma$ as a function of $\sigma$ by solving \eqref{v_ode} for $d=3$ and $\eta=0$.
 }
 \label{spike3dplot}
\end{figure}

\begin{figure}[tph]
 \includegraphics[width=8cm]{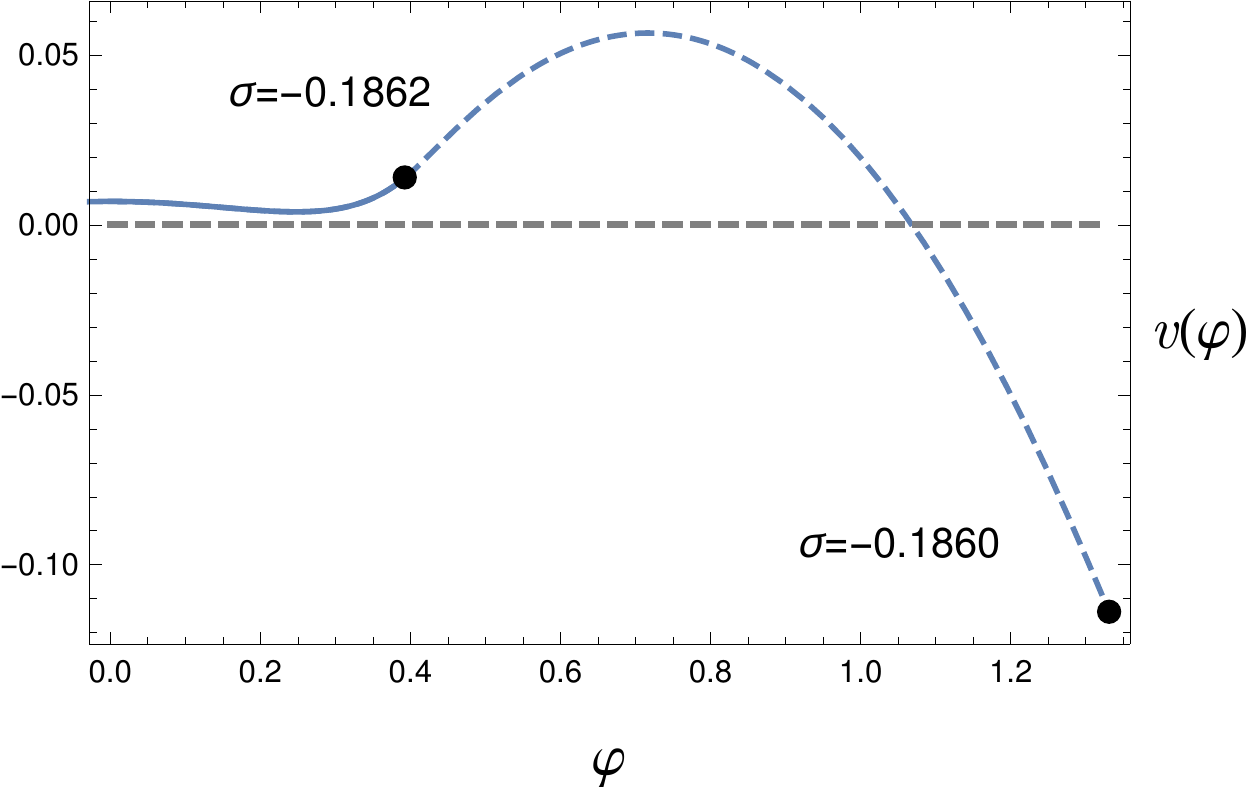}
 \caption{
 Almost critical solutions for the values $\sigma=-0.1860$ and $\sigma=-0.1862$.
 }
 \label{discontinuity}
\end{figure}

Critical values of the parameter $\sigma$ can be detected as discontinuities of either $\tilde{\varphi}_\sigma$ or its first derivative $\partial_\sigma \tilde{\varphi}_\sigma$.
In general, these are directly recognizable when plotting the function $\tilde{\varphi}_\sigma$ as either jumps or spikes.
In the universality classes of the scalar model they always appear as 
jumps of $\tilde{\varphi}_\sigma$.
Let us illustrate how the critical solutions are detected in practice by solving \eqref{v_ode} in the case of $d=3$.
We begin by numerically integrating \eqref{v_ode} with initial condition \eqref{v_zero} as a function of the parameter $\sigma$, with both formulas specialized to the case $d=3$ and $\eta=0$.
For almost all values of $\sigma$, the numerical integration comes to an end at a singularity $\tilde{\varphi}_\sigma$ defined by \eqref{singularity} and
whose value depends on $\sigma$ as shown in Fig.\ \ref{spike3dplot}.
The plot of $\tilde{\varphi}_\sigma$ reveals various salient features.
The singularity located at the point $\sigma=-1$ is due to our choice \eqref{v_zero} for the parametrization of the initial condition
and simply represents an infinite solution $v(\varphi)\to \infty$.
The singularity located in the center at $\sigma = 0$ is the Gaussian solution
which  trivially extends to all values of the field because the potential is identically constant $v(\varphi)={\rm const}$.
On the right of the plot the function $\tilde{\varphi}_\sigma$ slowly increases to infinity where another trivial solution,
representing an infinite mass limit is located.
The interesting feature of Fig.\ \ref{spike3dplot} is the singularity located on the left of the Gaussian one at $\sigma_{\rm cr}\approx-0.1861$,
which is characterized by a marked singular behavior of both $\tilde{\varphi}_\sigma$ (a jump) and its first derivative with respect to $\sigma$ (a spike).
We can try to interpret the meaning of this singularity by observing the behavior of the solution in the vicinity of the critical value $\sigma_{\rm cr}\approx-0.1861$
as shown in Fig.\ \ref{discontinuity}.
The solutions with $\sigma$ slightly on the left of $\sigma_{\rm cr}$ display a minimum
and end at the singularity \eqref{singularity} with positive values of $v(\tilde{\varphi}_{\sigma})$ and $v'(\tilde{\varphi}_\sigma)$.
Those with $\sigma$ slightly on the right of $\sigma_{\rm cr}$ end with negative values of $v(\tilde{\varphi}_\sigma)$ and $v'(\tilde{\varphi}_{\sigma})$.
The simultaneous change of signs in \eqref{singularity} is responsible for the sudden increase of $\tilde{\varphi}_\sigma$ while crossing $\sigma_{\rm}$ from left to right
as seen in Fig.\ \ref{spike3dplot}.

\begin{figure}[tph]
 \includegraphics[width=8cm]{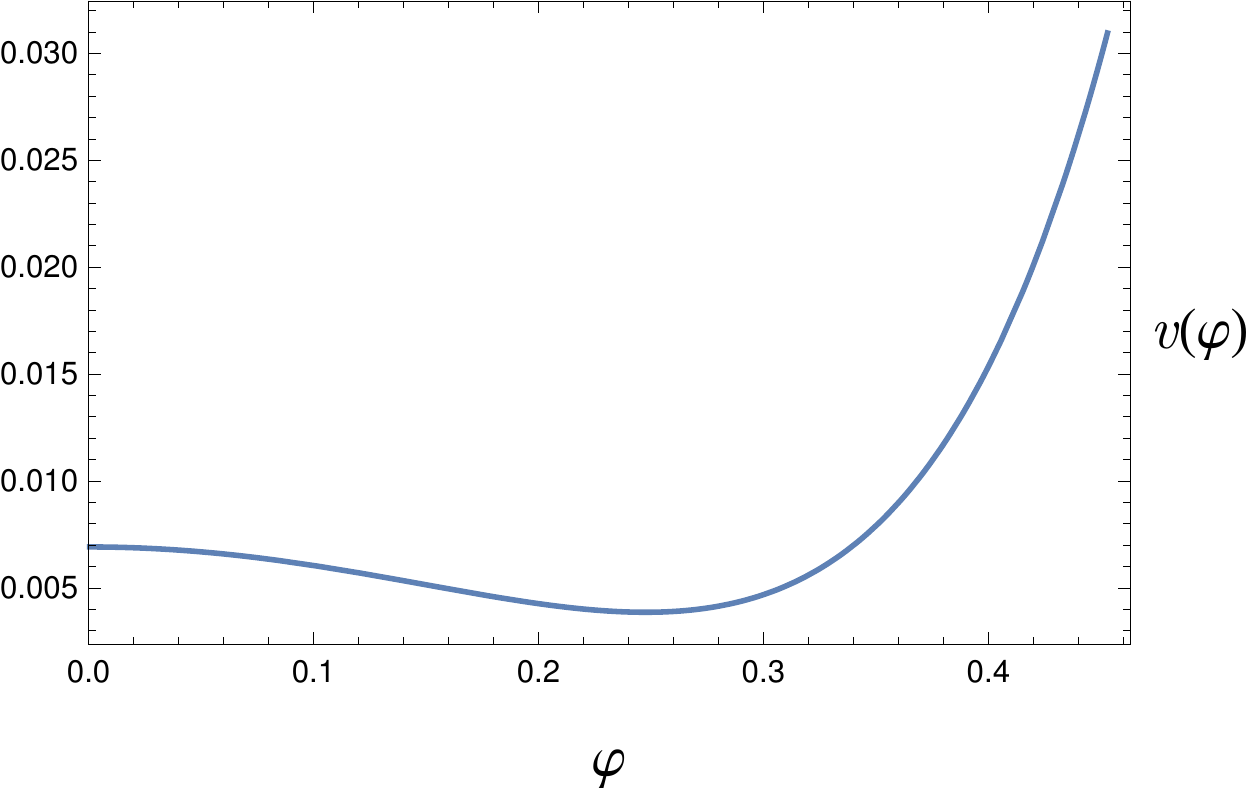}
 \caption{
 Critical solution obtained for the best determination of $ \sigma_{\rm cr}$.
 }
 \label{criticalsol}
\end{figure}

The left side of the non-trivial singularity in Fig.\ \ref{spike3dplot} 
displays solutions possessing a single non-trivial minimum
before terminating in the singularity. Therefore these are solutions with a non-trivial value of the order parameter $\varphi_0$ such that $v'(\varphi_0)=0$.
On the right of the singularity the solutions sharply bend to negative values and, when following $\sigma$ to zero, they continuously connect to the Gaussian solution.
It is thus tempting to define the critical solution as the one obtained from the limit $\sigma \to \sigma_{\rm cr}$ \emph{from the left}.
The value $\sigma_{\rm cr}$ has to be determined numerically and, within a ten digits precision, takes the value
$$ \sigma_{\rm cr}({\rm LPA}) = -0.1860662057\,. $$
A closer look at the numerically determined solution reveals the critical properties expected for the potential of the Ising-class in three dimension as shown in Fig.\ \ref{criticalsol}.
The critical potential in this figure has been calculated with 
the same precision by integrating the ODE \eqref{v_ode} using $\eta=0$ as an input.
We refer to the anomalous dimension appearing parametrically in \eqref{v_ode} as $\eta_{\rm in}$.
The consistency of the LPA can be further tested by using the solution itself to compute the anomalous dimension through the corresponding formula \eqref{eta}.
The new result for $\eta$ will be an output of the potential's determination 
and in general will be different from the input parameter.
We thus call the new anomalous dimension $\eta_{\rm out}$.
In the case of the potential of Fig.\ \ref{criticalsol} the minimum is located at $\varphi_0=0.2476$ and the use of \eqref{eta} yields $\eta_{\rm out}=0.1903$,
thus outlining that, strictly speaking, in the LPA the anomalous dimension is not determined.

The consistent determination of $\eta$ required by the framework of the LPA' can be achieved by constructing an algorithm that iteratively matches the input 
parameter for the anomalous dimension $\eta_{\rm in}$
with the output parameter $\eta_{\rm out}$ to a desired precision. The algorithm 
begins with a (reasonable) guess for $\eta_{\rm in}$ and
calculates the corresponding $v_{\sigma_{\rm cr}}(\varphi)$ which will 
be different from Fig.\ \ref{criticalsol}.
Then it computes $\eta_{\rm out}$ associated with the critical potential,
and finally iterates by choosing a new value of $\eta_{\rm in}$ closer 
to $\eta_{\rm out}$. Further details of the algorithms are discussed
in Appendix \ref{appendix_matching_algorithm}.
The result of the iteration is given with great accuracy 
in Subsect.\ \ref{subsection_ising3d}.

\subsection{Deformations of the simple scalar model}

Once both a critical value $\sigma_{cr}$ of $\sigma$ and the corresponding solution $v^*(\varphi)\equiv v_{\sigma_{\rm cr}}(\varphi)$ are identified,
it is possible to compute the critical properties of the solution in the 
form of critical exponents.
These are obtained by linearizing in the deviations from the critical solution, 
which means inserting the deformed solution
\begin{equation}
 v(\varphi)\to v^*(\varphi)+\epsilon\, \delta v(\varphi)\,\left(\frac{k}{k_0}\right)^{-\theta}\,
\end{equation}
into  \eqref{vdot}
and linearizing in the infinitesimal expansion parameter $\epsilon$. 
Here $k_0$ is a reference scale, and $\theta$ plays the role of the critical 
exponent of the deformation.
The linear order in $\epsilon$ gives an ODE for the fluctuation $\delta v(\varphi)$ that in canonical form reads
\begin{equation}
 \begin{split}\label{linearized_vdot}
  \delta v'' = &c_d^{-1} \frac{(d+2)}{d+2-\eta}\left[1+v^{*\prime\prime}(\varphi)
  \right]^2\,\times
  \\
  &\Bigl(
  (\theta-d)\delta v(\varphi)
  +\frac{1}{2}(d-2+\eta)\varphi\,\delta v'(\varphi)
  \Bigr)\,.
 \end{split}
\end{equation}
For later use, it proves convenient to perform a transformation that redefines the deformation as
\begin{equation}
\begin{split}\label{transformed_fluctuation}
 \delta u(\varphi) &\equiv \exp\left(-\alpha\int_0^\varphi {\rm d}y\,y 
 \big[1+v^{*\prime\prime}(y)\big]^2\right)\,\delta v(\varphi),\\
 &\alpha=\frac{d+2}{4c_d}\frac{d-2+\eta}{d+2-\eta}>0\,,
 \end{split}
\end{equation}
such that the linearized flow equation \eqref{linearized_vdot} does not 
contain the first derivative $\delta u'$.
The explicit form of the transformed ODE for $\delta u(\varphi)$ is 
involved and we found it not particularly instructive;
thus we do not write it down explicitly.

The second derivative $v^{*\prime\prime}(\varphi)$ in the linear fluctuation
equation \eqref{linearized_vdot} 
is an even function and we may assume that the deformations $\delta v(\varphi)$
are even or odd functions which means that they have the same parity 
as $v^*(\varphi)$ (parity-even)
or the opposite parity (parity-odd). Again, due to the linearity of \eqref{linearized_vdot}
we are free to choose the normalization of $\delta v(\varphi)$ at will.
This leaves us, modulo the global normalization, with two possible choices for the boundary conditions of the fluctuations at $\varphi=0$
\begin{eqnarray}
 \delta v(0)=1\,, & \quad\delta v'(0)=0\,, & \quad {\rm and}\\
 \delta v(0)=0\,, & \quad\delta v'(0)=1\,, &
\end{eqnarray}
for parity-even and parity-odd deformations, respectively. The same is of course true for the transformed deformations $\delta u(\varphi)$.
The spectrum of the critical exponents $\theta$ of the deformations is quantized when the deformations are required to be polynomially bounded,
thus adding a further condition at large values of the field.
We denote the quantized spectrum  of exponents as $\{ \theta_i^+,\theta_i^-\}$,
where $+/-$ label the even/odd parity of the deformation
and the index $i\in \mathbb{N}$ orders the exponents starting 
with the fluctuations with the highest positive critical exponent downward.

For the actual determination of the spectrum it is important to realize that,
in our setup, the solution $v^*(\varphi)$ is determined numerically and exists 
within a finite range only, $|\varphi|< \tilde{\varphi}_{\sigma_{\rm cr}}<\infty$.
Hence it is necessary to replace the boundary condition at infinity with 
appropriately chosen conditions at $\tilde{\varphi}_{\sigma_{\rm cr}}$
to reproduce the quantized spectrum of fluctuations. There are two 
natural ways to do so.

On the one hand the requirement that the fluctuations are subleading at 
infinity can be translated, within numerical accuracy,
to the requirement the fluctuations must tend to zero at the actual numerical boundary $\delta v(\tilde{\varphi}_{\sigma_{\rm cr}})=0$.
This can be achieved numerically with the help of the shooting
method.
The errors of introducing artificial boundary conditions for $\delta v(\varphi)$ 
can be controlled when one relaxes the condition by setting $\delta v(\tilde{\varphi}_{\sigma_{\rm cr}})=c$,
where $c$ is an arbitrary numerical constant. Typically, the critical 
exponents show a very mild dependence (of order $10^{-3}$) on $c$
even on the very large range $c\in [-10^3,10^3]$. This is due to the fact that the factor $[1+v^{*\prime\prime}(\varphi)]$ appearing in \eqref{linearized_vdot}
mitigates the effects of any boundary condition by suppressing it in the limit $\varphi\to\tilde{\varphi}_{\sigma_{\rm cr}}$. 
The dependence on the approximate boundary condition
can be relaxed even further by considering the transformed fluctuation \eqref{transformed_fluctuation},
for which it is natural to require $\delta u(\tilde{\varphi}_{\sigma_{\rm cr}})=0$.

On the other hand, since the numerical solution exists only in a 
finite range, it seems natural to impose periodic boundary conditions
for the even fluctuations and antiperiodic for the odd ones. We studied both cases by implementing the 
very efficient SLAC derivative method
for the transformed fluctuations defined in \eqref{transformed_fluctuation}
to compute the critical exponents.
In this case, the potential in the range $\varphi\in(-\tilde{\varphi}_{\sigma_{\rm cr}},\tilde{\varphi}_{\sigma_{\rm cr}})$ is discretized
over a equidistant grid such that the Schr\"odinger-type
differential equation for $\delta u(\varphi)$ turns into
a (large) matrix-equation in $\varphi$ space.
Most importantly, the derivative with respect to $\varphi$ is
discretized as non-local SLAC derivative. In the past this \emph{pseudo-spectral method}
based on trigonometric functions proved to be very efficient
to accurately calculate the low-lying energies of Schr\"odinger-type
operators. One may also estimate the characteristic width of an
eigenmode $\delta u(\varphi)$, which can be compared with the 
maximal field $\tilde{\varphi}_{\sigma_{\rm cr}}$
to decide whether the solution can be trusted and to what extent 
the spectrum is affected by finite size effects. 
Furthermore, the convergence of the spectrum can be checked by
increasing the number of grid points.
Not unexpectedly, the critical exponents obtained with the 
spectral method yield the most accurate results.
We refer to Appendix \ref{appendix_slac} for more details on the method.

\subsection{The Ising class in $d=3$}

\label{subsection_ising3d}

In this Subsection we provide tabulated results for the critical solution of \eqref{v_ode} in three dimensions
that corresponds to the Ising universality class. In each case we studied the critical value $\sigma_{\rm cr}$
with at least 10 digits precision, while the anomalous dimensions were determined with at least seven digits
\begin{equation}
 \begin{split}
  \sigma_{\rm cr}({\rm LPA\phantom{'}}) = -0.1860642495\,; & \\
  \sigma_{\rm cr}({\rm LPA'})= -0.1356610909\,, &\quad \eta\phantom{'}=0.1119795\,;\\
  \sigma_{\rm cr}({\rm LPA'})= -0.1657407105\,, &\quad \eta'=0.04427234\,.
 \end{split}
\end{equation}
The critical exponents were always determined with at least six digits precision.
In Table \ref{table_ising3d} the first three rows correspond to the LPA schemes described in the previous Sections,
including the LPA for which the input anomalous dimension $\eta_{\rm in}=0$,
the LPA' obtained by matching $\eta_{\rm in}$ with the anomalous dimension at the minimum using \eqref{eta}, and
the LPA' obtained by matching $\eta_{\rm in}$ with the anomalous dimension at the minimum using \eqref{goldstone_eta}.
The fourth row gives the result from a LPA in which the best known value of the anomalous dimension is taken from \cite{Pelissetto:2000ek}
as an input parameter, while the last three rows contain the best known results 
from various techniques \cite{Pelissetto:2000ek},
from lattice's finite-size scaling \cite{Hasenbusch:2011yya},
and from the conformal bootstrap approach \cite{Simmons-Duffin:2015qma}.
We provide values for the critical parameter $\sigma_{\rm cr}$, for both input and output anomalous dimensions, as well as for the critical exponent $\nu$ governing the scaling of the correlation length
which is defined as the inverse of the largest even critical exponent $\nu\equiv 1/\theta_1^+$.
We also give numerical estimations for the dimensionless coupling, defined as
\begin{equation}\label{critical_coupling3d}
 \begin{split}
  g_{\rm cr} &\equiv \frac{v^{(4)}(\varphi_0)}{6\,v''(\varphi_0)^{1/2}}\,.
 \end{split}
\end{equation}
Note that the definition \eqref{critical_coupling3d} refers to renormalized quantities and bears an intrinsic local potential scheme dependence.
The comparison of this critical coupling with analog definitions coming from a lattice's bare action (see for example \cite{Bosetti:2015lsa} and references therein)
might not be straightforward. The explicit relation between \eqref{critical_coupling3d}
and its lattice's counterpart could perhaps be computed along the lines of a non-perturbative generalization of \cite{Codello:2013bra}.
\begin{table}[h!]
\begin{tabular}{| l || r | r | r | r | r |}
 \hline
 & $\sigma_{\rm cr}$ & $g_{\rm cr}$ & $\eta_{\rm in}$ & $\eta_{\rm out}$ & $\nu$ 
 \\
 \hline
 LPA & -0.1861 & 16.1912 & 0 & 0.1903 & 0.6496
 \\
 \hline
 LPA' $\eta$ & -0.1357 & 9.6098 & 0.1120 & 0.1120 & 0.6453 
 \\
 \hline
 LPA' $\eta'$ & -0.1657 & 13.1693 & 0.0443 & 0.0443 & 0.6473
 \\
 \hline
 best $\eta$ & -0.1693 & 13.6776 & 0.0364 & 0.1626 & 0.6477 
 \\
 \hline
 \hline
 P.\&V.\ \cite{Pelissetto:2000ek} & -- & -- & 0.0364 & -- & 0.6301
 \\
 \hline
 Lattice \cite{Hasenbusch:2011yya} & -- & -- & 0.0363 & -- & 0.6300
 \\
 \hline
 Bootstrap \cite{Simmons-Duffin:2015qma} & -- & -- & 0.0363 & -- & 0.6300 
 \\
 \hline
\end{tabular}
\caption{Critical parameters for the Ising class in $d=3$.
The parameters are explained in the main text. The shooting and 
SLAC-method yield the same values for $\nu$.}
\label{table_ising3d}
\end{table}
All numbers of Tab.\ \ref{table_ising3d} are rounded to $10^{-4}$.

The non-smooth regulator \eqref{cutoff_optimized} might lead to problems
when it is used beyond the next to leading order of the derivative expansion \cite{Canet:2003qd}.
% \footnote{In particular,
% when using the optimized cutoff the flow of the (momentum space) two point function appearing in \eqref{extraction_scheme}
% is an analytic function of $p=\sqrt{p^2}$ rather than the expected $p^2$,
% thus making further derivatives with respect to $p^2$ ill-defined in the limit $p^2\to 0$.
% This does not happen when a smooth regulator is adopted.
% Our results are nevertheless well-defined within the LPA because the two point function in \eqref{extraction_scheme}
% is in fact free of the linear term in $p$.
% and therefore the expansion in $p$ can be interpreted as an expansion in $p^2$
% up to the second order.
% }
Thus, one may ascribe part of the error 
in determining the anomalous dimension to this particular
regulator instead of the LPA'-approximation. To investigate this
question we repeated the calculation with a power
law
regulator
\begin{equation}
 {\cal R}_k(q^2)=\frac{k^4}{q^2},\label{power_reg}
\end{equation}
and with the exponential regulators
\begin{equation}
{\cal R}_k(q^2)=\frac{aq^2}{e^{b y^c}-1},\quad y=\frac{q^2}{k^2}\label{exp_reg}
\end{equation}
which are compatible with the derivative expansion. The spike-plots
look very similar to those for the non-smooth regulator. The LPA'-results
with $\eta$ defined in (\ref{eta}) are
listed in Table \ref{table_ising3d_exponential}.
\begin{table}[h!]
\begin{tabular}{| l || r | r | r | r |}
\hline
cutoff& $\sigma_{\rm cr}$ & $g_{\rm cr}$ & $\eta$ & $\nu$ 
 \\
 \hline
 power law       & -0.3673 & 1.5528  & 0.1266 & 0.6505
 \\
 \hline
 exponential$_1$ & -0.3068 & 9.4133  & 0.1197 & 0.6462
 \\
 \hline
 exponential$_2$ & -0.2808 & 12.6742 & 0.0690 & 0.6530
 \\ 
 \hline
\end{tabular}
\caption{The same critical parameters obtained
with power law and exponential cutoffs. The results in
the row marked by exponential$_1$ are obtained with parameters
$(a,b,c)=(1,\log(2),1.44)$ (which are optimal for the  LPA) while
those in the line exponential$_2$ are obtained with $(a,b,c)=(1,1,1)$. The 
cutoffs are given in the main text.}
\label{table_ising3d_exponential}
\end{table}
We see that the anomalous dimension does not improve considerably compared
to the results for the non-smooth cutoff (\ref{cutoff_optimized}). It does improve a
bit for the particular choice $(a,b,c)=(1,1,1)$ of the parameters in the exponential
cutoff. But at the same time the result for the critical exponent
$\nu$ moves away from the accurate lattice-result $\nu=0.6300$ quoted in Table \ref{table_ising3d}. We conclude that the errors in determining the
anomalous dimension must be ascribed mainly to the LPA'-approximation.
Thus we proceed with the non-smooth regulator (\ref{cutoff_optimized})
for which the occurring momentum integrals can be calculating
explicitly.

Next we give in Table \ref{table_ising3d_even_exponents} the calculated values for the critical exponents corresponding to even deformations of the critical Ising solution.
For the various local potential schemes, we adopted both the shooting method which enforces the boundary condition $\delta v(\varphi)=0$ and the spectral SLAC-method.\footnote{
Here and for the rest of the paper we do not display the critical exponent $\theta^+_0$ corresponding to a constant deformation of the potential
that determines the scaling of the volume operator, since for \emph{all} critical models it is equal to the dimensionality of the system (which in this example is three).
We refer to Appendix \ref{appendix_sigma_dependence} for more details on $\theta^+_0$ and its peculiar behavior away from criticality.
}
\begin{table}[h!]
 \begin{tabular}{| l || r | r | r | r | r |}
 \hline
 Shooting: & $\theta^+_1$ & $\theta^+_2$ & $\theta^+_3$ & $\theta^+_4$ & $\theta^+_5$
 \\
 \hline
 LPA & 1.5395 & -0.6557 & -3.1800 & -5.9122 & -8.7961
 \\
 \hline
 LPA' $\eta$ & 1.5496 & -0.5323 & -2.9017 & -5.4604 & -8.1642
 \\
 \hline
 LPA' $\eta'$ & 1.5448 & -0.6076 & -3.0729 & -5.7415 & -8.5789
 \\
 \hline
 best $\eta$ & 1.5440 & -0.6162 & -3.0919 & -5.7699 & -8.5993
 \\
 \hline
 \hline
  SLAC: & $\theta^+_1$ & $\theta^+_2$ & $\theta^+_3$ & $\theta^+_4$ & $\theta^+_5$
 \\
 \hline
 LPA & 1.5395 & -0.6557 & -3.1800 & -5.9122 & -8.7961 
 \\
 \hline
 LPA' $\eta$ & 1.5496  & -0.5323 & -2.9017 & -5.4598 & -8.1585 
 \\
 \hline
 LPA' $\eta'$ & 1.5448 & -0.6076 & -3.0727 & -5.7384 & -8.5513
 \\
 \hline
 best $\eta$ & 1.5440 & -0.6163 & -3.0920 & -5.7698 & -8.5956
 \\
 \hline
\end{tabular}
\caption{Critical exponents of the even deformations for the Ising class in $d=3$.
}
\label{table_ising3d_even_exponents}
\end{table}

Finally, we give in Table \ref{table_ising3d_odd_exponents} the numerical values for the critical exponents corresponding to odd deformations of the Ising scaling solution,
again displaying the results of both shooting and spectral SLAC-method.
\begin{table}[h!]
 \begin{tabular}{| l || r | r | r | r | r |}
 \hline
 Shooting: & $\theta^-_1$ & $\theta^-_2$ & $\theta^-_3$ & $\theta^-_4$ & $\theta^-_5$
 \\
 \hline
 LPA & 2.5000 & 0.5000 & -1.8867 & -4.5244 & -7.3376
 \\
 \hline
 LPA' $\eta$ & 2.4440 & 0.5560 & -1.6892 & -4.1635 & -6.8188
 \\
 \hline
 LPA' $\eta'$ & 2.4779 & 0.5221 & -1.8103 & -4.3854 & -7.1387
 \\
 \hline
 best $\eta$ & 2.4818 & 0.5182 & -1.8240 & -4.4097 & -7.1665
 \\
 \hline
 \hline
 SLAC: & $\theta^-_1$ & $\theta^-_2$ & $\theta^-_3$ & $\theta^-_4$ & $\theta^-_5$
 \\
 \hline
 LPA & 2.5000 & 0.5000 & -1.8867 & -4.5244 & -7.3377
 \\
 \hline
 LPA' $\eta$ & 2.4440 & 0.5560 & -1.6891 & -4.1608 & -6.7933
 \\
 \hline
 LPA' $\eta'$ & 2.4779 & 0.5221 & -1.8102 & -4.3845 & -7.1288
 \\
 \hline
 best $\eta$ & 2.4818 & 0.5182 & -1.8240 & -4.4097 & -7.1665
 \\
 \hline
\end{tabular}
\caption{Critical exponents of the odd deformations for the Ising class in $d=3$.}
\label{table_ising3d_odd_exponents}
\end{table}

As expected, in all cases we detect one relevant deformation
having the same even parity as the critical solution. Two more relevant
deformations show up in the odd sector and can be related to the anomalous 
dimension $\eta_{\rm in}$ which governs the scaling part as explained in Appendix \ref{appendix1}.
In particular, the first two exponents of the odd sector have the numerical values
$\theta^-_1\simeq (5-\eta_{\rm in})/2$ and $\theta^-_2\simeq (1+\eta_{\rm in})/2$
consistently with the results of Appendix \ref{appendix1}.

Our estimates for the critical exponent $\nu$ deviate at most by
$3$ percent from the most accurate values compiled in \cite{Pelissetto:2000ek}.
The agreement is, however, not as excellent for the anomalous dimension
and, in particular, for the LPA' scheme based on the definition \eqref{eta},
which differs from the best known value by a factor three.
This discrepancy is a known limitation of the local potential schemes and
could be eased by improving the LPA to a higher order derivative expansion \cite{Morris:1994ie,Morris:1996kn,Morris:1994ki}.
We discuss this point further in Sect.\ \ref{section_discussion}.

For completeness we also list the critical exponents for the even and odd 
deformations as obtained with the power law regulator (\ref{power_reg})
and the exponential regulator (\ref{exp_reg}) in Table \ref{table_ising3d_exponential_exponents}. 
The parameters of the exponential regulator are the same as
in Table \ref{table_ising3d_exponential},
and again the anomalous dimension is determined with the formula (\ref{eta}). 
The exponents are consistent with those
calculated with the non-smooth regulators and listed in
Tables
\ref{table_ising3d_even_exponents} and 
\ref{table_ising3d_odd_exponents}.
\begin{table}[h!]
 \begin{tabular}{| l || r | r | r | r | r |}
 \hline
 cutoff & $\theta^+_1$ & $\theta^+_2$ & $\theta^+_3$ & $\theta^+_4$ & $\theta^+_5$
 \\
 \hline
 power law & 1.5373 & -0.4954 & -2.7695 & -5.1950 & -7.7301
 \\
 \hline
 exponential$_1$ & 1.5475 & -0.5207 & -2.8689 & -5.4011 & -8.0726
 \\ 
 \hline
 exponential$_2$ & 1.5313 & -0.5671 & -2.9482 & -5.5109 & -8.2175
 \\
 \hline
 \hline
 cutoff & $\theta^-_1$ & $\theta^-_2$ & $\theta^-_3$ & $\theta^-_4$ & $\theta^-_5$
 \\
 \hline
 power law & 2.4367 & 0.5633 & -1.6095 & -3.9665 & -6.4505
 \\
 \hline
 exponential$_1$ & 2.4402 & 0.5598 & -1.6676 & -4.1155 & -6.7210
 \\
 \hline 
 exponential$_2$ & 2.4655 & 0.5345 & -1.7301 & -4.2101 & -6.8468
 \\
 \hline
 \end{tabular}
\caption{Critical exponents of the even and odd deformations for the Ising 
class in $d=3$ and calculated with a power-law and two exponential
regulators. The parameters are the same as in Table
\ref{table_ising3d_exponential}.}
\label{table_ising3d_exponential_exponents}
\end{table}

\subsection{The Ising class in $d=2$}

\label{subsection_ising2d}

\begin{figure}[tph]
 \includegraphics[width=8cm]{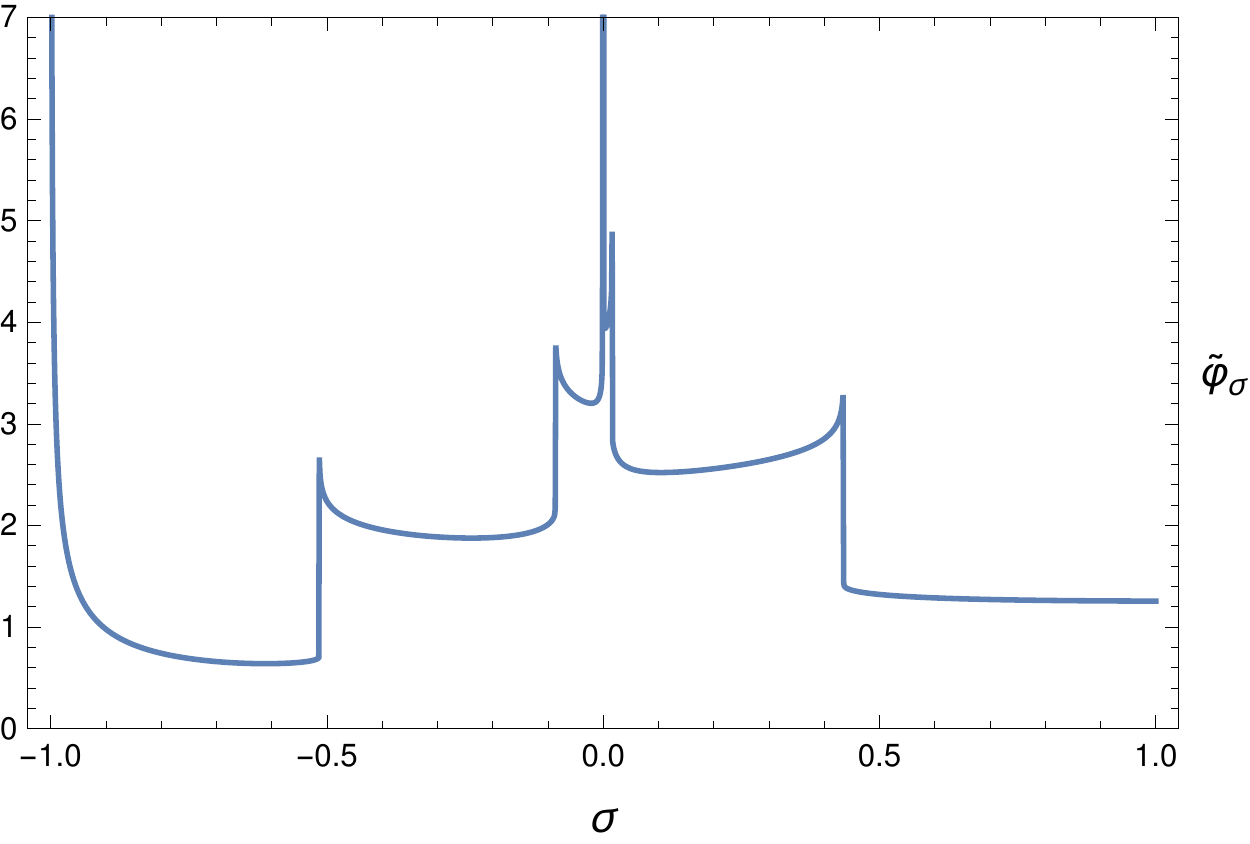}
 \caption{
 Location of the singularities $\tilde{\varphi}_\sigma$ as a function of $\sigma$ by solving \eqref{v_ode} for $d=2$ and $\eta=1/4$.
 }
 \label{spike2dplot}
\end{figure}

In two dimensions and for \emph{zero anomalous dimension} the 
fixed point equation \eqref{v_ode} can be integrated without 
ever encountering a singularity for all initial conditions.
The resulting solutions have been argued to belong to the Sine-Gordon 
universality class \cite{Codello:2012sc}, meaning that the LPA is not capable
of detecting the two-dimensional critical solutions known 
as minimal models.
To circumvent this deficiency one may continue \eqref{v_ode} to a dimensionality 
slightly above two $d \gtrsim 2$,
or allow for a non-trivial anomalous dimension $\eta_{\rm in}$. 
In both cases, the effective dimensionality of the field $(d-2+\eta_{\rm in})/2$,
which vanished in LPA, becomes positive.
Fixing $d=2$ and using a non-zero $\eta_{\rm in}$, it is possible to compute the function $\tilde{\varphi}_\sigma$ as in the previous section.
With decreasing $\eta_{\rm in}$ the corresponding plot of $\sigma\to\tilde{\varphi}_\sigma$ 
shows an increasing number of singularities,
which we identify with the well-known minimal conformal models in two dimensions.
Fig.\ \ref{spike2dplot} shows the spike-plot of $\tilde{\varphi}_\sigma$ 
for $\eta_\mathrm{in}=1/4$.
It reveals that four critical solutions are already present
for this choice of $\eta$. In the limit of effective field's dimension $(d-2+\eta_{\rm in})/2\to 0$,
solutions appear spawning from the Gaussian one at $\sigma=0$ and alternating from left to right.
A closer inspection reveals that the leftmost non-trivial singularity corresponds to an Ising class,
the rightmost to a tri-Ising class, the second from the left to a quadri-Ising and the second from the right to a penta-Ising.
It appears that the heights of the singularities, that is their approximated domain of existence,
are ordered like the inverse of the central charge of the corresponding minimal model.

We now concentrate our attention on the LPA' of the two-dimensional Ising class;
that is, we isolate the leftmost singularity of Fig.\ \ref{spike2dplot} and match the input and output anomalous dimensions.
Like in the previous Subsections, this can be done by either using $\eta$ or the Goldstone's $\eta'$, resulting in two different sets of critical parameters
\begin{equation}
 \begin{split}
  \sigma_{\rm cr}({\rm LPA'})=-0.3583283424\,, &\quad \eta\phantom{'}=0.436348\,;\\
  \sigma_{\rm cr}({\rm LPA'})=-0.5291046049\,, &\quad \eta'=0.233594\,.
 \end{split}
\end{equation}
Analogously to \eqref{critical_coupling3d} we define the two-dimensional 
critical coupling as the ratio
\begin{equation}\label{critical_coupling2d}
 \begin{split}
  g_{\rm cr} &\equiv \frac{v^{(4)}(\varphi_0)}{6\,v''(\varphi_0)}\,,
 \end{split}
\end{equation}
which corresponds to the canonically dimensionless ratio $g_{\rm cr}=\lambda/m^2$ of 
the quartic coupling and squared mass in two dimensions.

In Tab.\ \ref{table_ising2d} we listed the critical parameters 
for the Ising class in two dimensions. The first two rows contain the LPA'-values 
obtained with the anomalous dimensions $\eta$ and $\eta'$.
In the third row $\eta_{\rm in}$ is set equal to the exact value
taken from CFT, and
finally, the last row contains the exact results coming from CFT.
\begin{table}[h!]
 \begin{tabular}{| l || r | r | r | r | r | r |}
 \hline
 & $\sigma_{\rm cr}$ & $g_{\rm cr}$ & $\eta_{\rm in}$ & $\eta_{\rm out}$ 
 & $\nu_\mathrm{shoot}$ &$\nu_{\rm slac}$
 \\
 \hline
 LPA' $\eta$ & -0.3583 & 13.0397 & 0.4363 & 0.4363 & 1.0550 & 1.0551
 \\
 \hline
 LPA' $\eta'$ & -0.5291 & 55.5004 & 0.2336 & 0.2336 & 1.3265 & 1.3264
 \\
 \hline
 Exact $\eta$ &  -0.5143 & 47.9118 & 0.2500 & 0.7477 & 1.2871 & 1.2870
 \\
 \hline
 CFT & -- &-- & 0.2500 & -- & 1.0000 & 1.0000
 \\
 \hline
\end{tabular}
\caption{Critical parameters for the Ising class in $d=2$.}
\label{table_ising2d}
\end{table}

In Tab.s \ref{table_ising2d_even_exponents} and \ref{table_ising2d_odd_exponents} 
we listed the critical exponents for parity even and odd deformations
of the critical solutions. As before, we provide both the results obtained 
with the shooting and the spectral SLAC-methods.
\begin{table}[h!]
 \begin{tabular}{| l || r | r | r | r | r |}
 \hline
 Shooting: & $\theta^+_1$ & $\theta^+_2$ & $\theta^+_3$ & $\theta^+_4$ & $\theta^+_5$
 \\
 \hline
 LPA' $\eta$ & 0.9478 & -0.6761 & -2.6194 & -4.7373 & -6.9765
 \\
 \hline
 LPA' $\eta'$ & 0.7539 & -0.8935 & -2.9866 & -5.3007 & -7.7637
 \\
 \hline
 Exact $\eta$ & 0.7770 & -0.8741 & -2.9597 & -5.2613 & -7.7111
 \\
 \hline
 \hline
 SLAC: & $\theta^+_1$ & $\theta^+_2$ & $\theta^+_3$ & $\theta^+_4$ & $\theta^+_5$
 \\
 \hline
 LPA' $\eta$ & 0.9478 & -0.6761 & -2.6193 & -4.7347 & -6.9938
 \\
 \hline
 LPA' $\eta'$ & 0.7539 & -0.8935 & -2.9863 & -5.2960 & -7.7125
 \\
 \hline
 Exact $\eta$ & 0.7770 & -0.8741 & -2.9593 & -5.2547 & -7.6424
 \\
 \hline
\end{tabular}
\caption{Critical exponents of the even deformations for the Ising class in $d=2$.}
\label{table_ising2d_even_exponents}
\end{table}
\begin{table}[h!]
 \begin{tabular}{| l || r | r | r | r | r |}
 \hline
 Shooting: & $\theta^-_1$ & $\theta^-_2$ & $\theta^-_3$ & $\theta^-_4$ & $\theta^-_5$
 \\
 \hline
 LPA' $\eta$ & 1.7818 & 0.2182 & -1.6189 & -3.6605 & -5.8438
 \\
 \hline
 LPA' $\eta'$ & 1.8832 & 0.1168 & -1.8903 & -4.1204 & -6.5150
 \\
 \hline
 Exact $\eta$ & 1.8750 & 0.1250 & -1.8703 & -4.0880 & -6.4689
 \\
 \hline
 \hline
 SLAC: & $\theta^-_1$ & $\theta^-_2$ & $\theta^-_3$ & $\theta^-_4$ & $\theta^-_5$
 \\
 \hline
 LPA' $\eta$ & 1.7818 & 0.2182 & -1.6189 & -3.6606 & -5.8452
 \\
 \hline
 LPA' $\eta'$ & 1.8832 & 0.1168 & -1.8903 & -4.1206 & -6.5168
 \\
 \hline
 Exact $\eta$ & 1.8750 & 0.1250 & -1.8703 & -4.0880 & -6.4696
 \\
 \hline
\end{tabular}
\caption{Critical exponents of the odd deformations for the Ising class in $d=2$.}
\label{table_ising2d_odd_exponents}
\end{table}

The estimates for the critical exponent of the correlation length in Tab.\ \ref{table_ising2d} deviate
from the exact values determined within CFT by $6\,\%-30\,\%$ with the standard LPA' scheme giving the best agreement.
On the other hand and similarly to Subsetc.\ \ref{subsection_ising3d}, the LPA' scheme provides the least accurate
value for the anomalous dimension. We refer to the discussion in Sect.\ \ref{section_discussion} for more details concerning this point.

\subsection{The Tri-Ising class in $d=2$}

\label{subsection_tri_ising3d}

In this Subsection we briefly summarize our findings about the rightmost 
singularity appearing in the spike-plot associated to the tricritical Ising class
in Fig.\ \ref{spike2dplot}.
The properties of this class can be investigated by adopting the 
same techniques that were used for the Ising class.
Again, the critical parameters can be determined by either using $\eta$ or the Goldstone's $\eta'$ for the matching of $\eta_{\rm in}$ and $\eta_{\rm out}$,
thus resulting in two different sets of critical parameters
\begin{equation}
 \begin{split}
  \sigma_{\rm cr}({\rm LPA'})=0.2597409360\,, &\quad \eta\phantom{'}=0.311938\,;\\
  \sigma_{\rm cr}({\rm LPA'})=2.9267843351\,, &\quad \eta'=0.0686649\,.
 \end{split}
\end{equation}
We also supplement the critical coupling $g_\mathrm{ cr}$ in
(\ref{critical_coupling2d}) by
\begin{equation}\label{critical_couplings_tricritical_ising}
g_{6,{\rm cr}} \equiv \frac{v^{(6)}(\varphi_0)}{120\, v''(\varphi_0)}\,,
\end{equation}
which corresponds to the dimensionless ratio $g_{6,{\rm cr}}=\lambda_6/m^2$ for the expansion
\begin{equation}
 \begin{split}\label{bare_potential2}
  v(\varphi) &\simeq \frac{m^2}{2} \varphi^2 + \frac{\lambda}{4}\varphi^4 + \frac{\lambda_6}{6}\varphi^6+\dots\,,
 \end{split}
\end{equation}
accommodating a higher criticality for the tricritical Ising class.

The critical parameters are determined within the various local potential schemes in Tab.\ \ref{table_triising2d}.
In Tab.s \ref{table_triising2d_even_exponents} and \ref{table_triising2d_odd_exponents} we give the critical exponents for parity even and odd deformations. 
We provide only the results obtained with the spectral SLAC-method,
but we also compare with the exact results coming from CFT.
\begin{table}[h!]
 \begin{tabular}{| l || r | r | r | r | r | r |}
 \hline
 & $\sigma_{\rm cr}$ & $g_{\rm cr}$& $g_{6,{\rm cr}}$ & $\eta_{\rm in}$ & $\eta_{\rm out}$  &$\nu$
 \\
 \hline
 LPA' $\eta$ & 0.2597 & 8.4410 & 23.1728 & 0.3119 & 0.3119 & 0.6001
 \\
 \hline
 LPA' $\eta'$ & 2.9268 & 182.135 & 19210 & 0.0687 & 0.0687 & 0.5239
 \\
 \hline
 Exact $\eta$ &  1.0639 & 38.8756& 747.039 & 0.1500 & 0.7256 & 0.5498
 \\
 \hline
 CFT & -- & -- & --& 0.1500 & -- & 0.5560 
 \\
 \hline
\end{tabular}
\caption{Critical parameters for the tricritical Ising class in $d=2$.}
\label{table_triising2d}
\end{table}
\begin{table}[h!]
 \begin{tabular}{| l || r | r | r | r | r |}
 \hline
 SLAC: & $\theta^+_1$ & $\theta^+_2$ & $\theta^+_3$ & $\theta^+_4$ & $\theta^+_5$
 \\
 \hline
 LPA' $\eta$ & 1.6665 & 0.6836 & -0.4510 & -1.7920 & -3.2840
 \\
 \hline
 LPA' $\eta'$ & 1.9087 & 0.3596 & -0.7899 & -2.1364 & -3.7899
 \\
 \hline
 Exact $\eta$ & 1.8187 & 0.5501 & -0.6248 & -2.0554 & -3.6878
 \\
 \hline
\end{tabular}
\caption{Critical exponents of the even deformations for the tricritical Ising class in $d=2$.}
\label{table_triising2d_even_exponents}
\end{table}
\begin{table}[h!]
 \begin{tabular}{| l || r | r | r | r | r |}
 \hline
  SLAC: & $\theta^-_1$ & $\theta^-_2$ & $\theta^-_3$ & $\theta^-_4$ & $\theta^-_5$
 \\
 \hline
 LPA' $\eta$  & 1.8440 & 1.1560 & 0.1560 & -1.1008 & -2.5218
 \\
 \hline
 LPA' $\eta'$ & 1.9657 & 0.6890 & 0.03434 & -0.7899 & -2.1364 \\
 \hline
 Exact $\eta$ & 1.9250 & 0.9715 & 0.07500 & -1.3157 & -2.8562
 \\
 \hline
\end{tabular}
\caption{Critical exponents of the odd deformations for the tricritical Ising class in $d=2$.}
\label{table_triising2d_odd_exponents}
\end{table}

It is worth noting that our definition of the critical couplings \eqref{critical_couplings_tricritical_ising}
displays a wide range of values. This is especially true for the quantity $g_{6,{\rm cr}}$ which varies
over several orders of magnitude between the various LPA schemes. The numerical 
values of the parameters $g_{\rm cr}$ and $g_{6,{\rm cr}}$
can be used in \eqref{bare_potential2} to obtain an approximate form of the local potential solutions.
Then a decreasing anomalous dimension in Tab.\ \ref{table_triising2d}
leads to a solution with a more pronounced minimum,
which means that the coefficients in the Taylor-expansion have large numerical values.

\section{The ${\cal N}=1$ Wess-Zumino model}

\label{section_wz_model}

The most straightforward supersymmetric extension of the scalar field theory 
is the supersymmetric ${\cal N}=1$ Wess-Zumino model (WZM).
Since in $3$-dimensional \emph{Euclidean} space there are no Majorana-spinors
we formulate the flow equations in Minkowski space.
Then Majorana-spinors exist in three and two dimensions and 
have two components in both Minkowski spaces.
Hence an interpolation of the flow equations from three to two dimensions
should be possible.
Both in three and two dimensions
the field content of the model can be characterized through the superfield
\begin{equation}
 \Phi = \varphi + (\bar{\theta}\psi) + \frac{1}{2} (\bar{\theta}\theta) F\,,
\end{equation}
in which $\varphi$ is a real scalar field, $\psi$ a Majorana spinor 
and $F$ a real auxiliary field.
We defined the Grassmann variable $\theta$ and its conjugate $\bar{\theta}$ which will play the role of coordinates in superspace.

Following previous works \cite{Synatschke:2009da,Synatschke:2009nm,Synatschke:2010ub} we formulate the LPA by using a manifestly supersymmetric 
truncation for the effective average action in superspace,
\begin{equation}\label{wz_effective_average_action}
 \Gamma_k=\int{\rm d}^dx\! \int {\rm d} \theta\, {\rm d} \bar \theta \Bigl(-\frac{1}{2}Z_k \Phi K \Phi+2 W_k(\Phi)\Bigr)\,,
\end{equation}
in which we introduced the second order superderivative operator $K$.
This truncation generalizes \eqref{scalar_effective_average_action} 
by introducing the scale-dependent effective superpotential $W_k(\Phi)$.
Integrating explicitly over the Grassmann variables \eqref{wz_effective_average_action}
can be expressed in terms of the component fields
\begin{equation}
\begin{split}
 \Gamma_k=\int{\rm d}^dx &\Bigl(
 \frac{Z_k}{2}\partial_\mu\varphi\partial^\mu \varphi-\frac{i}{2}Z_k\bar\psi\slashed\partial\psi+\frac{Z_k}{2}F^2
 \\
 &-\frac{1}{2}W_k''(\varphi)(\bar\psi\psi)+W_k'(\varphi)F
 \Bigr)\,.
\end{split}
\end{equation}
It is instructive to eliminate the auxiliary field $F$ through its 
algebraic equation of motion to obtain the so-called on-shell
effective average action
\begin{equation}
\begin{split}\label{wz_effective_average_action_on_shell}
 \Gamma_k^{\rm on \, shell}
 =\int{\rm d}^dx &\Bigl(
 \frac{Z_k}{2}\partial_\mu\varphi\partial^\mu \varphi+\frac{W_k'^2(\varphi)}{2Z_k}
 \\
 &-\frac{i}{2}Z_k\bar\psi\slashed\partial\psi-\frac{1}{2}W_k''(\varphi)(\bar\psi\psi)\Bigl)\,.
\end{split}
\end{equation}
A direct comparison of the first line with \eqref{scalar_effective_average_action} shows that, on-shell, the scalar sector of the WZM is equivalent
to a scalar field theory controlled by a potential $V_k(\varphi)=W_k'^2(\varphi)/2Z_k$ and coupled to the Majorana field $\psi$
with a particular Yukawa interaction whose form is constrained by supersymmetry.

It is possible to construct an infrared cutoff like \eqref{cutoff} which is 
quadratic in the field multiplet $\{\varphi,\psi,F\}$ and that preserves 
supersymmetry, thus giving rise to a renormalization group flow that is fully 
covariant with respect to supersymmetry.
We refer to \cite{Synatschke:2009da,Synatschke:2009nm,Synatschke:2010ub} 
for more details on the construction of 
a manifest supersymmetric flow, while we give the simplest implementation
that (after Wick-rotation to Euclidean space) in momentum space is
\begin{equation}\label{susy_cutoff_action}
 \begin{split}
  \Delta S_k = \frac{Z_k}{2} \int {\rm d}^d q \Bigl(&
  \varphi_q {\cal R}_k(q^2)\varphi_{-q}
  +F_q \frac{{\cal R}_k(q^2)}{q^2} F_{-q}
  \\
  &
  -\frac{i}{2} \bar\psi_q\,\slashed q\, \frac{{\cal R}_k(q^2)}{q^2}\psi_{-q}
  \Bigr)\,,
 \end{split}
\end{equation}
which depends on a single cutoff kernel ${\cal R}_k(q^2)$ carrying the 
scheme dependence of the flow.
Notice that a manifestly supersymmetric RG-flow with a regulator quadratic
in the fields can only be formulated in the off-shell formulation.
The auxiliary field can only be eliminated after the complete FRG-flow has 
been determined. 
Nonetheless, we still refer to 
$V_k(\varphi)=W_k'^2(\varphi)/2 Z_k$ as the scale-dependent effective potential 
of the WZ-model,
remembering that in the limit $k\to 0$ it does coincide with the effective potential for the scalar mode $\varphi$.

As for the simple scalar model it is possible to explicitly evaluate the RG flow of the model within the given truncation
once the cutoff kernel ${\cal R}_k(q^2)$ is explicitly chosen.
A simple family of kernels that mimics \eqref{cutoff_optimized} is
\begin{equation}\label{susy_cutoff}
 {\cal R}^n_k(q^2) = q^2\big(\left(k/\left|q\right|\right)^n-1\big)\,\theta(k^2-q^2)\,,
\end{equation}
which is parametrized by the exponent $n$.
The choice $n=2$ leads to a finite flow both in two and three dimensions,
while the choice $n=1$, which coincides with \eqref{cutoff_optimized}, yields
a divergent flow in two dimensions. In complete analogy to the scalar 
field theory, we introduce the dimensionless renormalized field $\overline{\varphi}_{\rm R}\equiv Z_k^{-1/2}k^{(2-d)/2}\varphi $
that is the natural argument of the dimensionless renormalized superpotential
\begin{equation}
 w_k(\overline{\varphi}_{\rm R}) \equiv k^{-d/2} W_k(\varphi)/Z_k\,,
\end{equation}
with anomalous dimension $\eta \equiv - k\partial_k Z_k/Z_k$.
Once again, for notational simplicity we omit the labels in the following.
In the evaluation of the Wick-rotated flow \eqref{erge} 
within the truncation \eqref{wz_effective_average_action}
the dependence on the space-dimension is easily seen in momentum
space. This way one arrives at a $d$-dependent flow for the superpotential
which serves as a natural analytic continuation of the RG flow of the model 
to dimensions $2\leq d \leq 3$.

The flow is computed in practice by projecting \eqref{erge} to the simple field configuration $\Phi =\varphi = {\rm const.}$ Using
\eqref{susy_cutoff} for $n=2$ we obtain the system
\begin{eqnarray}\label{susy_wdot}
%  \begin{split}
  && k\partial_k w(\varphi) = 
  {\cal S}_{w,\eta}[w,w';\varphi]+{\cal F}_{w,\eta}[w'']\,,\nonumber
  \\
  &&{\cal S}_{w,\eta}[w,w';\varphi] \equiv (1-d)w(\varphi)+\frac{d-2+\eta}{2}\varphi 
  w'(\varphi)\,,
  \\
  &&{\cal F}_{w,\eta}[w''] \equiv
  -\frac{c_d}{2}\Bigl(
   (2-\eta){\cal A}_{1,0}[w'']+\frac{d\eta}{(d+2)}{\cal A}_{1,1}[w'']
   \Bigr) w''
   \,,\nonumber
%  \end{split}
\end{eqnarray}
where we maintained a compact notation by defining
\begin{equation}\label{susy_hyper_int}
\begin{split}
 {\cal A}_{\,l,m}[w'']&\equiv {_2}F_1\left(l,m+d/2,m+1+d/2,-w''^2\right)\\
 &=\left(m+\frac{d}{2}\right)\int_0^1 \mathrm{d}t\,\frac{t^{m+d/2-1}}{(1+tw''^2)^ {l}}
 \end{split}
\end{equation}
which depends on the Gauss hypergeometric function ${_2}F_1$. The integral
representation shows explicitly that for non-negative $l$ and $m$ the function 
$\mathcal{A}_{l,m}[w'']$ is a positive function on the real axis
which satisfies
\begin{equation}\label{susy_A_prop}
  {\cal A}_{\,l,m}[-w'']={\cal A}_{\,l,m}[w''],\quad  {\cal A}_{\,l,m}[0]=1\,.
\end{equation}
It follows that ${\cal F}_{w,\eta}$ is an odd function of $w''$ which is 
negative for positive arguments. Its negative slope at the origin is
\begin{equation}\label{susy_slope_F}
{\cal F}'_{w,\eta}[0]=-c_d\left(1-\frac{\eta}{d+2}\right)\,.
\end{equation}
In an off-shell formulation of supersymmetric 
field theories there is an efficient way to extract the anomalous dimension, see
\cite{Synatschke:2009da,Synatschke:2009nm,Heilmann:2014iga} for details.
For consistency the anomalous dimension must be evaluated at a particular
field configuration,
which we choose to be the order parameter $\varphi_0$.
Introducing the dimensionless renormalized on-shell potential
\begin{equation}
 v(\varphi)\equiv 
 \frac{w'(\varphi)^2}{2}\,,
\end{equation}
the order parameter is defined as its (nontrivial) minimum through the relation $v'(\varphi_0)=0$.
Using the compact notations $w''_0=w''(\varphi_0)$ and 
$w'''_0=w'''(\varphi_0)$, the anomalous dimension can be expressed implicitly as
\begin{equation}\label{susy_eta}
 \begin{split}
  \eta =c_d\Bigl[\,& (2-\eta) {\cal A}_{\,3,0}[w''_0]+\frac{d\eta}{d+2}{\cal A}_{\,3,1}[w''_0] 
  \\
  &
  -\frac{d(2-\eta)}{d+2}{\cal A}_{\,3,1}[w''_0]\,{w''_0}^2 w'''_0
  \\
  &-
  \frac{d\eta}{d+4}{\cal A}_{\,3,2}[w''_0]\,{w''_0}^2 \,\Bigr] {w'''_0}^2.
 \end{split}
\end{equation}
In the limit $d\downarrow 2$ the equations \eqref{susy_wdot} and \eqref{susy_eta} 
can be seen to reproduce  the results in \cite{Synatschke:2009da,Synatschke:2009nm}.
Analogous results are obtained for the choice $n=1$ in \eqref{susy_cutoff}
that are finite above $2$ dimensions, and in the limit $d\uparrow 3$ reproduce the results of \cite{Synatschke:2010ub}.
Our computation with the cutoff $n=2$ offers a consistent analytic continuation of the flow of the superpotential which interpolates between the known results for $d=2,3$.
The results obtained with the $n=1$ cutoff will only be used 
in $3$ dimensions to determine, to some extent, the cutoff independence of the methods.

To conclude this Section, let us note that, in contrast to the non-supersymmetric case, 
the anomalous dimension \eqref{susy_eta}
has, for some models, a nonzero value when evaluated at the configuration $\varphi_0=0$.
Let us thus define a further anomalous dimension
\begin{equation}\label{susy_eta0}
 \eta_0 = \left.\eta\right|_{\varphi_0=0}\,.
\end{equation}
For all the critical solutions that we are interested in, the location $\varphi_0=0$ is either a minimum or a local maximum,
but never the global minimum which is needed in the LPA'.
This anomalous dimension is thus perturbative, in the sense that it 
is not obtained by an expansion about the true vacuum state.
Nevertheless, for the WZ-model it is possible to investigate a new 
local potential scheme, which we call LPA$^0$, that uses $\eta_0$ as an 
input for the integration of \eqref{susy_wp_ode}.
This partly perturbative scheme lies in between the LPA and the LPA'.
\footnote{
In passing we note that this scheme was originally dubbed as LPA'
in \cite{Synatschke:2009da,Synatschke:2009nm},
where a different definition for the local potential approximation
for the WZ-model has been used.
}

\subsection{Scaling solutions for the WZM}

It seems difficult to \emph{explicitly} rearrange 
the ODE corresponding to the stationary solutions 
of the highly non-linear flow equation \eqref{susy_wdot} 
in a standard form $w''=\dots$ as we did for the scalar field
in \eqref{v_ode}. 
Rather than finding stationary solutions for $w(\varphi)$ via \eqref{susy_wdot}, we decided to investigate the stationary solutions of the flow equation 
for $w'(\varphi)$, which can be cast in standard form.\footnote{
It is actually possible to numerically solve \eqref{susy_wdot}
by applying an implicit solver combined with a residual method such as Newton's.
This however comes at the expense of the solution's accuracy.
We explicitly checked that the results reported below can be reproduced
by the sole use of \eqref{susy_wdot}.
}
To obtain this
equation we differentiate  \eqref{susy_wdot} with respect to $\varphi$
and solve the fixed-point equation $k \partial_k w'(\varphi)=0$ 
for $w'''(\varphi)$ with the result
\begin{equation}\label{susy_wp_ode}
w'''(\varphi)=\frac{1}{{\cal F}_{w,\eta}'(w'')}\Bigl(\frac{d-\eta}{2}w'-\frac{d-2+\eta}{2}\varphi w''\Bigr)\,.
\end{equation}
However, the asymptotic structure of stationary solutions
is more directly (and more comfortably) accessible from \eqref{susy_wdot} itself.
For large values of $w''$ the function ${\cal F}_{w,\eta}$ appearing on the right hand side of \eqref{susy_wdot} has the asymptotic expansion
\begin{equation}\label{susy_asymptotic}
%\begin{align}
{\cal F}_{w,\eta}=\begin{dcases}-\frac{c_d}{2w''}\left(\eta +(2-\eta)\frac{d}{d-2}\right)
, & 2<d<4\,;\\
-\frac{c_2}{2 w''} \left(\eta +(2-\eta)  \log \big(w''^{\,2}\big)\right), & d=2\,,
\end{dcases}
%\end{align}
\end{equation}
which implies, like in the scalar field theory, a divergent second derivative of the superpotential when the scaling part of the equation is zero.
Neglecting the leading logs and overall factors
\begin{equation}\label{susy_singularity1}
w''(\varphi) \sim \frac{1}{{\cal S}_{w,\eta}}\,, \quad {\rm when}\quad {\cal S}_{w,\eta}\sim 0\,.
\end{equation}

A closer inspection of \eqref{susy_wp_ode} reveals that the ODE might possess
a singularity that is met whenever ${\cal F}'_{w,\eta}=0$.
This is indeed true because the function ${\cal F}_{w,\eta}$ 
intersects the $w''$-axis at the origin with negative slope, see
\eqref{susy_slope_F}. Since in addition ${\cal F}_{w,\eta}[w'']$
tends to zero for large arguments, see \eqref{susy_asymptotic},
we conclude that there must exist a minimum $w''^*$ of ${\cal F}_{w,\eta}$ at 
which ${\cal F}'_{w,\eta}$ vanishes.
When integrating \eqref{susy_wp_ode} we expect to find a singular $w'''$ 
when $w''$ approaches $w''^*$. In other words, for generic
initial conditions we find
\begin{equation}\label{susy_singularity2}
w'''(\varphi) \sim \infty\,, \quad {\rm when}\quad {\cal F}'_{w,\eta}(w'')\to 0\,,
\end{equation}
therefore there exists a singular value of the field $\tilde{\varphi}$ for which
\begin{equation}\label{susy_phimax}
w''(\varphi) \to w''^*\,, \quad {\rm when}\quad \varphi\to \tilde{\varphi}\,.
\end{equation}
A comparison of the definition \eqref{susy_phimax} with the analog definition for the simple scalar model \eqref{singularity}
reveals that the integrations of the stationary ODEs meet the singularities in two rather different ways.
It turns out that the singularity \eqref{susy_phimax} can in fact be crossed if
\begin{equation}
\frac{\rm d}{{\rm d}\varphi}{\cal S}_{w,\eta} = -\frac{d-\eta}{2}w'+\frac{d-2+\eta}{2}\varphi w''=0
\end{equation}
at $\varphi=\tilde{\varphi}$, which will turn out to be true only for all the critical solutions that we will investigate.
When this condition is met at a critical point, the third derivative of the superpotential does not diverge
according to \eqref{susy_singularity2} and the solution acquires the desired \emph{global} nature.
For more details on this extension we refer to Appendix \ref{appendix_crossing}.

The behavior of the singularities must be studied as a function of the 
initial conditions of the fixed point equation \eqref{susy_wp_ode}.
Our only constraint is that the on-shell potential must be $\mathbb{Z}_2$-invariant.
Recalling that ${\cal F}'_{w,\eta}$ in \eqref{susy_wp_ode} is an even 
function  of $w''$ it is easy to see that the function $w'(\varphi)$ must 
either be even or odd in $\varphi$.
\emph{If $w'(\varphi)$ is even} then $w''(0)=0$ and we parametrize 
the initial condition with $w'''(0) =\zeta_-$
so that
\begin{equation}\label{wp_bc_odd_w}
\begin{split}
\begin{dcases}
 w'(0)&=-\frac{2 \,c_d \, (d-\eta +2)}{(d+2) (d-\eta )}\,\zeta_-\,,\\
 w''(0) &=0\,,
\end{dcases}
\end{split}
\end{equation}
which thus serve as boundary conditions for parity odd superpotentials.
\emph{If instead $w'(\varphi)$ is odd} then $w'(0)=0$
and we choose $w''(0) = \zeta_+$ so that
\begin{equation}\label{wp_bc_even_w}
\begin{split}
\begin{dcases}
 w'(0)  &=0\,,\\
 w''(0) &= \zeta_+\,,
\end{dcases}
\end{split}
\end{equation}
which are the appropriate boundary conditions for parity even superpotentials.

The situation is different from the previously considered scalar field theory 
in which one parameter $\sigma$ in \eqref{v_zero} was sufficient
to parametrize all possible critical models. 
In the supersymmetric case the solutions fall into two distinct
classes characterized by different initial conditions
\eqref{wp_bc_odd_w} and \eqref{wp_bc_even_w}
with corresponding parameters $\zeta_{\pm}>0$.
For the ease of numerical comparison between the scalar-model LPA and 
the WZ-model LPA,
we can however provide, using the on-shell dimensionless potential $v(\varphi)=w'^2(\varphi)/2$,
the value of $\sigma\equiv v''(0)$ which is defined in analogy to the previous Sections. We obtain
\begin{equation}\label{susy_sigma}
\begin{split}
\begin{dcases}
 \sigma(\zeta_-)&= -\frac{2 \,c_d \, (d-\eta +2)}{(d+2) (d-\eta )}\,\zeta^2_-<0 \,, \quad {\rm or}\\
 \sigma(\zeta_+)  &= \zeta_+^2>0 \,,
\end{dcases}
\end{split}
\end{equation}
for even or odd superpotentials respectively.
This shows that the first class will contain solutions with an even number 
of minima as it happens for $\sigma<0$,
while the second class contains solutions with an odd number of minima.
We dedicate the next Subsections to the critical models that can be 
detected by plotting the maximum value of the field \eqref{susy_phimax}
as a function of $\zeta_{\pm}$.

\subsection{The SUSY-Ising class in $d=3$}

\label{subsection_susy_ising3d}

\begin{figure}[tph]
 \includegraphics[width=8cm]{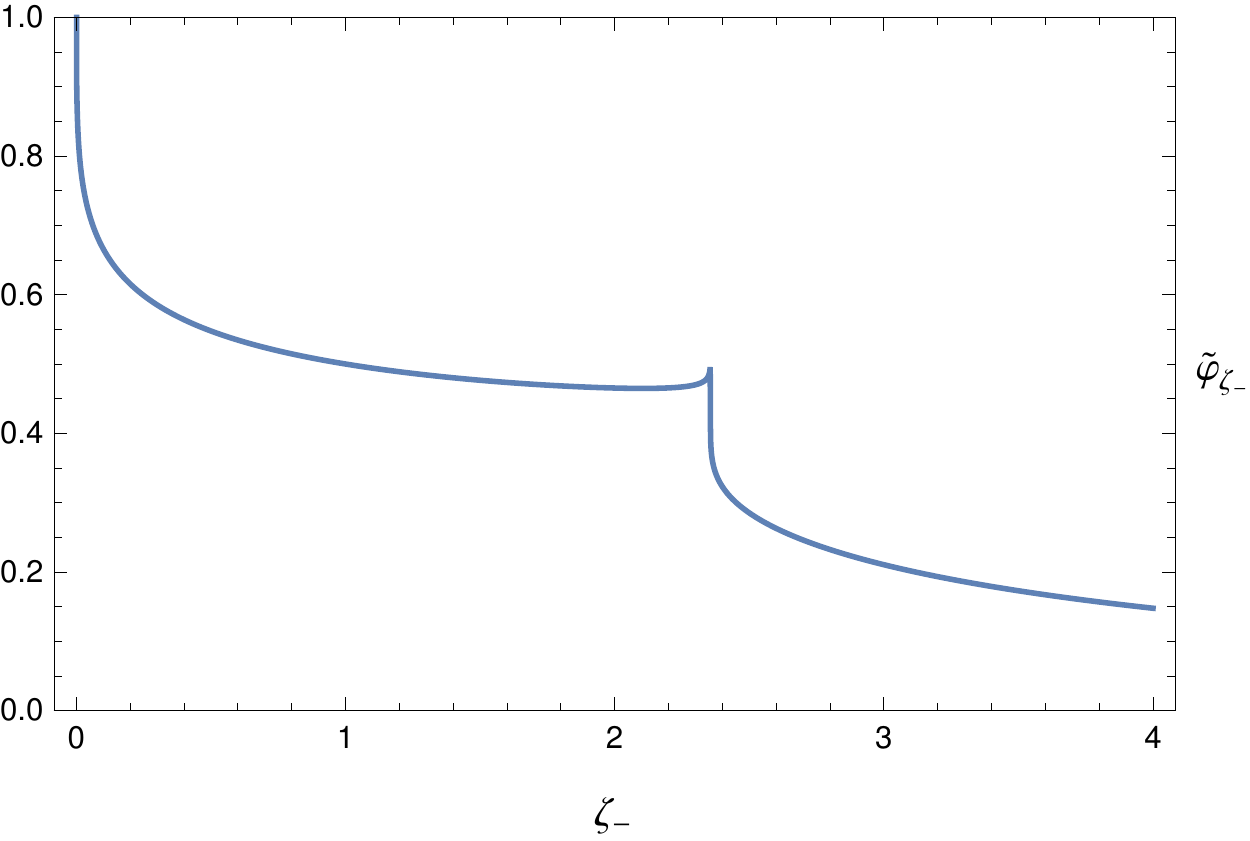}
 \caption{
 Location of the singularity of $\tilde{\varphi}_{\zeta_-}$ as a function of $\zeta_-$ by solving \eqref{susy_wp_ode} (LPA$_0$) for $d=3$ and $\eta=\eta_0$ using the $n=1$ cutoff.
 }
 \label{susy_spike3dplot}
\end{figure}

The supersymmetric extension of the Ising class in $3$ dimensions
is obtained
by integrating the fixed point equation \eqref{susy_wp_ode} with
the boundary conditions \eqref{wp_bc_odd_w} for $d=3$.
It is sufficient to study the range of initial conditions for which $\zeta_{-}>0$, since the negative $\zeta_{-}$ can be reached by mirror symmetry.
Within this range, all numerical solutions of the fixed point equation 
end at a singularity $\tilde{\varphi}_{\zeta_-}$ obeying \eqref{susy_phimax} 
such that $w''$ diverges.
The plot of $\tilde{\varphi}_{\zeta_-}$ is given in Fig.\ \ref{susy_spike3dplot} and should remind the reader of the negative $\sigma$ range of Fig.\ \ref{spike3dplot},
as in fact the parameters $\sigma$ and $\zeta_-$ can be explicitly related through \eqref{susy_sigma}.
Even though the jump at the critical value $\zeta_{-,{\rm cr}}$
is not as pronounced as in the non-supersymmetric case, 
the plot of Fig.\ \ref{susy_spike3dplot} clearly displays
a singularity at which the maximal field $\tilde{\varphi}_{\zeta_-}$ 
is discontinuous and its first derivative diverges.
The singularity is not always as pronounced, and in particular 
it is absent in the LPA scheme with the $n=2$ cutoff.
We therefore decided to display the plot using the LPA${}^0$ 
scheme defined by \eqref{susy_eta0},
being the closest analogue to the LPA. This, in particular, implies that, by varying $\zeta_-$ in Fig.\ \ref{susy_spike3dplot}
both the value of $w'(0)$ \emph{and} of $\eta_0$ vary.

A closer inspection of the solutions in the vicinity of the singularity reveals that
the solutions with $\zeta_-$ larger than the critical value have an Ising-like potential,
while those with smaller $\zeta_-$ correspond to solutions which bend down and, with decreasing value of $\zeta_{-,{\rm cr}}$, an additional minimum develops.
This is in complete analogy with the situation described in Subsect.\  \ref{subsection_scaling_solutions}
for the Ising-class, especially if we recall
that $\sigma\propto -\zeta^2_-$. We thus investigate the 
supersymmetric Ising solution as the limit $\zeta_-\downarrow \zeta_{-,{\rm cr}}$ 
\emph{from above}.

Similarly as for the scalar field theory, we now apply 
the numerical algorithms discussed in the previous section to 
investigate the critical properties of the supersymmetric 
fixed point solutions. In 
Table \ref{table_susy_ising3d} we provide
the critical value for $\zeta_-$,
the corresponding critical value $\sigma_{\rm cr}$ (using \eqref{susy_sigma}),
the value of the critical coupling (which is defined analogously to \eqref{critical_coupling3d} by using the on-shell potential),
the anomalous dimension
and the critical exponent of the correlation length.
These numerical estimates are given for the choices $n=1,2$ of the cutoff \eqref{susy_cutoff}
and for the available schemes of LPA${}^0$ and LPA'. The critical values of the parameters are:
\begin{equation}
 \begin{split}
\zeta_-({\rm LPA}^0{}_{n=1})=2.023871922\,,& \quad \eta\phantom{'}=0.188003\,;\\
\zeta_-({\rm LPA}'{}_{n=1})=2.133306022\,, &\quad \eta\phantom{'}=0.173624\,;\\
\zeta_-({\rm LPA}^0{}_{n=2})=2.355854024\,,& \quad \eta\phantom{'}=0.180673\,;\\
\zeta_-({\rm LPA}'{}_{n=2})=2.479354615\,, &\quad \eta\phantom{'}=0.167018\,.
 \end{split}
\end{equation}
\begin{table}
\begin{tabular}{| l || r | r | r | r | r | r |}
 \hline
 & $\zeta_{-,{\rm cr}}$ & $\sigma_{\rm cr}$ & $g_{\rm cr}$ & $\eta_{\rm in}$ & $\eta_{\rm out}$ & $\nu$
 \\
 \hline
 LPA$^0{}_{n=1}$ & 2.0239 & -0.0692 & 13.5783 & 0.1880 & 0.1599 & 0.7112
 \\
 \hline
 LPA'$_{n=1}$ & 2.1333 & -0.0384 & 10.4531 & 0.1736 & 0.1736 & 0.7076
 \\
 \hline
 LPA$^0{}_{n=2}$ & 2.3558 & -0.0641 & 10.1464 & 0.1807 & 0.1531 & 0.7094
 \\
 \hline
 LPA'$_{n=2}$ & 2.4794 & -0.0708 & 11.0943 & 0.1670 & 0.1670 & 0.7060
 \\
 \hline
\end{tabular}
\caption{Critical parameters for the supersymmetric Ising class in $d=3$.}
\label{table_susy_ising3d}
\end{table}

In Tables \ref{table_susy_ising3d_even_exponents} and \ref{table_susy_ising3d_odd_exponents}
we provide our best estimates for the critical exponents of the 
parity-odd and parity-even deformations of the superpotential, 
which correspond to deformations of the on-shell
potential of \emph{opposite} parity.
\begin{table}[h!]
\begin{tabular}{| l || r | r | r | r | r |}
\hline
 Shooting: & $\theta_1^+$ & $\theta_2^+$ & $\theta_3^+$ & $\theta_4^+$ & $\theta_5^+$
 \\ 
 \hline
 LPA$^0{}_{n=1}$ & 1.4060 & -0.3510 & -2.5715 & -5.1730 & -8.1200
 \\
 \hline
 LPA'$_{n=1}$ & 1.4132 & -0.3824 & -2.6813 & -5.4004 & -8.5042
 \\
 \hline
 LPA$^0{}_{n=2}$ & 1.4097 & -0.3500 & -2.5281 & -5.0357 & -7.8328
 \\
 \hline
 LPA'$_{n=2}$ & 1.4165 & -0.3773 & -2.6200 & -5.2216 & -8.1418
 \\
 \hline
 \hline
 SLAC: & $\theta_1^+$ & $\theta_2^+$ & $\theta_3^+$ & $\theta_4^+$ & $\theta_5^+$
 \\
 \hline
 LPA$^0{}_{n=1}$ & 1.4060 & -0.3510 & -2.5715 & -5.1730 & -8.1200
 \\
 \hline
 LPA'$_{n=1}$ & 1.4130 & -0.3824 & -2.6813 & -5.4004 & -8.5043
 \\
 \hline
 LPA$^0{}_{n=2}$ & 1.4097 & -0.3500 & -2.5281 & -5.0357 & -7.8328
 \\
 \hline
 LPA'$_{n=2}$ & 1.4165 & -0.3773 & -2.6200 & -5.2216 & -8.1418
 \\
 \hline
\end{tabular}
\caption{Critical exponents of the even deformations for the on-shell potential of the supersymmetric Ising class in $d=3$.}
\label{table_susy_ising3d_even_exponents}
\end{table}
\begin{table}[h!]
\begin{tabular}{| l || r | r | r | r | r |}
 \hline
 Shooting: & $\theta^-_1$ & $\theta^-_2$ & $\theta^-_3$ & $\theta^-_4$ & $\theta^-_5$
 \\
 \hline
 LPA$^0{}_{n=1}$ & 0.5940 & -1.4102 & -3.8274 & -6.6048 & -9.7168
 \\
 \hline
 LPA'$_{n=1}$ & 0.5868 & -1.4759 & -3.9911 & -6.9056 & -10.1945
 \\
 \hline
 LPA$^0{}_{n=2}$ & 0.5903 & -1.3941 & -3.7438 & -6.3997 & -9.3328
 \\
 \hline
 LPA'$_{n=2}$ & 0.5835 & -1.4500 & -3.8781 & -6.6435 & -9.7144
 \\
 \hline
 \hline
 SLAC: & $\theta^-_1$ & $\theta^-_2$ & $\theta^-_3$ & $\theta^-_4$ & $\theta^-_5$
 \\
 \hline
 LPA$^0{}_{n=1}$ & 0.5940 & -1.4102 & -3.8274 & -6.6048 & -9.7168
 \\
 \hline
 LPA'$_{n=1}$ & 0.5868 & -1.4760 & -3.9916 & -6.9056 & -10.1945
 \\
 \hline
 LPA$^0{}_{n=2}$ & 0.5903 & -1.3941 & -3.7438 & -6.3997 & -9.3328
 \\
 \hline
 LPA'$_{n=2}$ & 0.5835 & -1.4500 & -3.8791 & -6.6435 & -9.7144
 \\
 \hline
\end{tabular}
\caption{Critical exponents of the odd deformations for the on-shell potential of the supersymmetric Ising class in $d=3$.}
\label{table_susy_ising3d_odd_exponents}
\end{table}

The first exponent of Tab.\ \ref{table_susy_ising3d_even_exponents} corresponds to the deformation
$\delta w(\varphi) \propto \varphi$, which, according to the discussion given in Appendix \ref{appendix1}, has
critical exponent $(d-\eta)/2$, and is also related to the critical exponent of the correlation length by the relation $\nu=1/\theta^+_1$.
We can thus test that the relation
\begin{equation}\label{superscaling}
 \theta^+_1=\frac{1}{\nu} = \frac{d-\eta}{2}\,,
\end{equation}
known as \emph{superscaling relation} \cite{Heilmann:2014iga}, is satisfied within our accuracy.
The first critical exponent of Tab. \ref{table_susy_ising3d_odd_exponents} corresponds instead to
the deformation $\delta w(\varphi)\sim w'(\varphi)$, which is also discussed in Appendix \ref{appendix1}
and whose critical exponent is related to the anomalous dimension by the formula
\begin{equation}\label{magnetic}
 \theta^-_1 = \frac{d-2+\eta}{2}\,.
\end{equation}
This relation is perfectly satisfied within our accuracy.

\subsection{The SUSY-Ising class in $d=2$}
\label{subsection_susy_ising2d}

\begin{figure}[tph]
 \includegraphics[width=8cm]{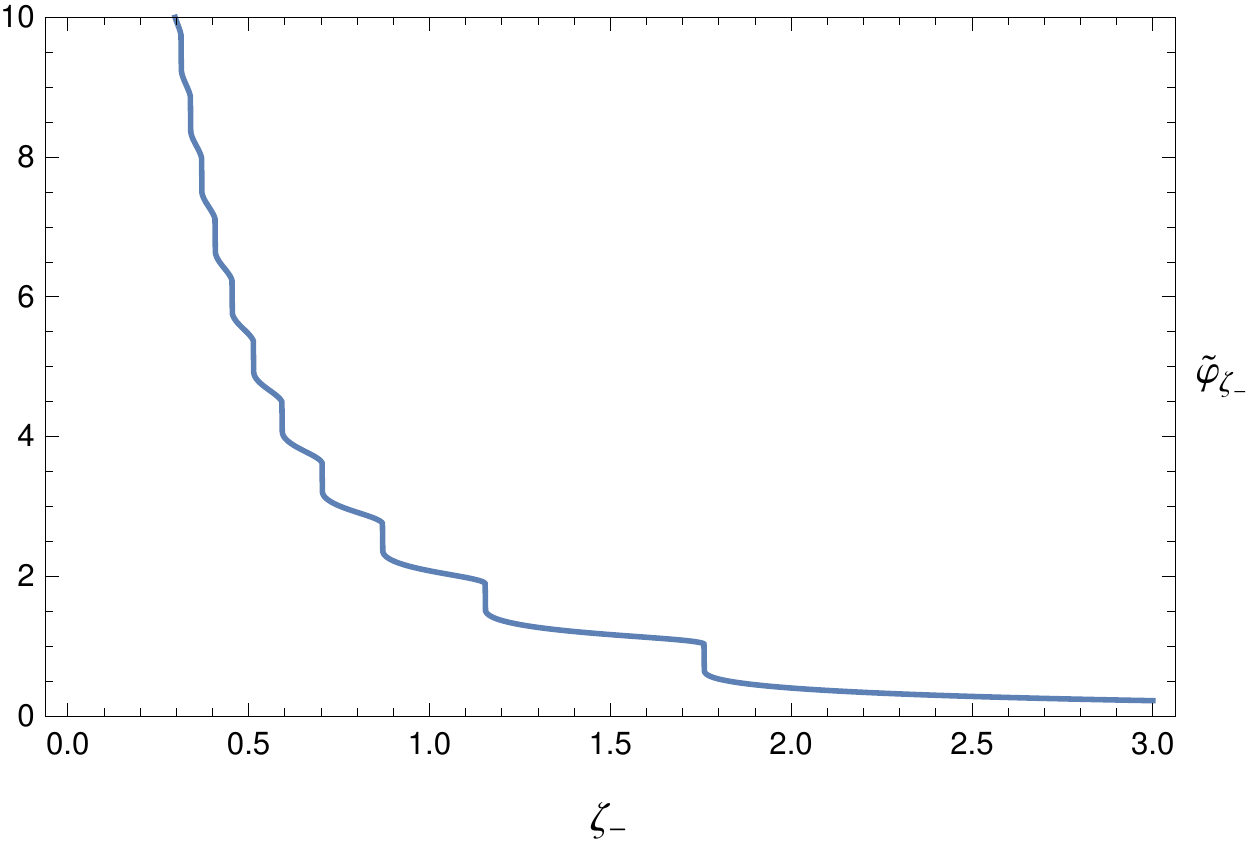}
 \caption{
 Location of the singularities of $\tilde{\varphi}_{\zeta_-}$ as a function of $\zeta_-$ by solving \eqref{susy_wp_ode} for $d=2$ and $\eta=\eta_0$ (LPA$_0$) using the $n=1$ cutoff.
 The rightmost singularity corresponds to the Ising model, and then towards the origin all \emph{even} solutions with higher criticalities appear.
 }
 \label{susy_spike2dplot}
\end{figure}

A supersymmetric extension of 
the Ising class in $2$ dimensions is obtained by
integrating the fixed point equation \eqref{susy_wp_ode} with
the boundary conditions \eqref{wp_bc_odd_w} for $d=2$.
The plot of the endpoints of integrations is given within the LPA$^0$ 
scheme in Fig.\ \ref{susy_spike2dplot} and shows that, as expected, 
many critical points emerge in the two dimensional limit.
A direct inspection of the solutions and the relation \eqref{susy_sigma} 
reveal that these singularities correspond to even critical solutions of 
higher  criticality.
Apparently, the plot in Fig.\ \ref{susy_spike2dplot} looks 
qualitatively different from our previous plots such as 
in Fig.\ \ref{susy_spike3dplot}:
It does not show spike-like singularities, but rather what appear to be 
bends or flexes. It can, however, be checked that at each singular point the function $\tilde{\varphi}_{\zeta_-}$
is indeed discontinuos with a diverging first derivative. The only difference from all the previous plots is that the first derivative of $\tilde{\varphi}_{\zeta_-}$ 
has the same sign on both sides of the singularity.

Approaching the rightmost singularity from
the right 
we obtain a supersymmetric extension of the Ising class.
The critical values for the two available schemes are
\begin{equation}
 \begin{split}
\zeta_-({\rm LPA}^0{}_{n=2})=1.7593594599\,,& \quad \eta\phantom{'}=0.438619\,;\\
\zeta_-({\rm LPA}'{}_{n=2})=1.9477913122\,, &\quad \eta\phantom{'}=0.397108\,.
 \end{split}
\end{equation}
The numbers for all critical parameters are listed in Tab.\ \ref{table_susy_ising2d}.
As before
we also provide the numbers for the critical exponents of even and odd deformations of the critical solution in Tab.s \ref{table_susy_ising2d_even_exponents} and \ref{table_susy_ising2d_odd_exponents}
respectively.
\begin{table}[h!]
\begin{tabular}{| l || r | r | r | r | r | r |}
 \hline
 & $\zeta_{-,{\rm cr}}$ & $\sigma_{\rm cr}$ & $g_{\rm cr}$ & $\eta_{\rm in}$ & $\eta_{\rm out}$ & $\nu$
 \\
 \hline
 LPA$^0{}_{n=2}$ & 1.7594 & -0.2809 & 9.0576 & 0.4386 & 0.3386 & 1.2809
 \\
 \hline
 LPA'${}_{n=2}$ & 1.9478 & -0.3393 & 11.6385 & 0.3971 & 0.3971 & 1.2478
 \\
 \hline
\end{tabular}
\caption{Critical parameters for the supersymmetric Ising class in $d=2$.}
\label{table_susy_ising2d}
\end{table}
\begin{table}[h!]
\begin{tabular}{| l || r | r | r | r | r |}
 \hline
Shooting: & $\theta_1^+$ & $\theta_2^+$ & $\theta_3^+$ & $\theta_4^+$ & $\theta_5^+$
 \\
 \hline
 LPA$^0{}_{n=2}$ & 0.7807 & -0.4385 & -2.2244 & -4.5811 & -7.5113
 \\
 \hline
 LPA'${}_{n=2}$ & 0.8014 & -0.5132 & -2.5460 & -5.3257 & -8.8585
 \\
 \hline
 \hline
 SLAC: & $\theta_1^+$ & $\theta_2^+$ & $\theta_3^+$ & $\theta_4^+$ & $\theta_5^+$
 \\
 \hline
 LPA$^0{}_{n=2}$ & 0.7807 & -0.4383 & -2.2224 & -4.5711 & -7.4777
 \\
 \hline
 LPA'${}_{n=2}$ & 0.8014 & -0.5125 & -2.5416 & -5.3095 & -8.8144
 \\
 \hline
\end{tabular}
\caption{Critical exponents of the even deformations for the on-shell potential of the supersymmetric Ising class in $d=2$.}
\label{table_susy_ising2d_even_exponents}
\end{table}
\begin{table}[h!]
\begin{tabular}{| l || r | r | r | r | r |}
 \hline
Shooting: & $\theta^-_1$ & $\theta^-_2$ & $\theta^-_3$ & $\theta^-_4$ & $\theta^-_5$
 \\
 \hline
 LPA$^0{}_{n=2}$ & 0.2193 & -1.2610 & -3.3316 & -5.9739 & -9.1951
 \\
 \hline
 LPA'${}_{n=2}$ & 0.1984 & -1.4384 & -3.8423 & -6.9975 & -10.9101
 \\
 \hline
 \hline
 SLAC: & $\theta^-_1$ & $\theta^-_2$ & $\theta^-_3$ & $\theta^-_4$ & $\theta^-_5$
 \\
 \hline
 LPA$^0{}_{n=2}$ & 0.2193 & -1.2602 & -3.3268 & -5.9549 & -9.1393
 \\
 \hline
 LPA'${}_{n=2}$ & 0.1986 & -1.4365 & -3.8335 & -6.9700 & -10.8432
 \\
 \hline
\end{tabular}
\caption{Critical exponents of the odd deformations for the on-shell potential of the supersymmetric Ising class in $d=2$.}
\label{table_susy_ising2d_odd_exponents}
\end{table}
The first critical exponents of both Tab.s \ref{table_susy_ising2d_even_exponents} and \ref{table_susy_ising2d_odd_exponents}
are related to the anomalous dimension through the relations \eqref{superscaling} and \eqref{magnetic} specialized to $d=2$.
Both are verified well within our accuracy, giving yet another check of the superscaling relation.

\subsection{The SUSY-Tri-Ising class in $d=2$}
\label{subsection_susy_tri_ising2d}

\begin{figure}[tph]
 \includegraphics[width=8cm]{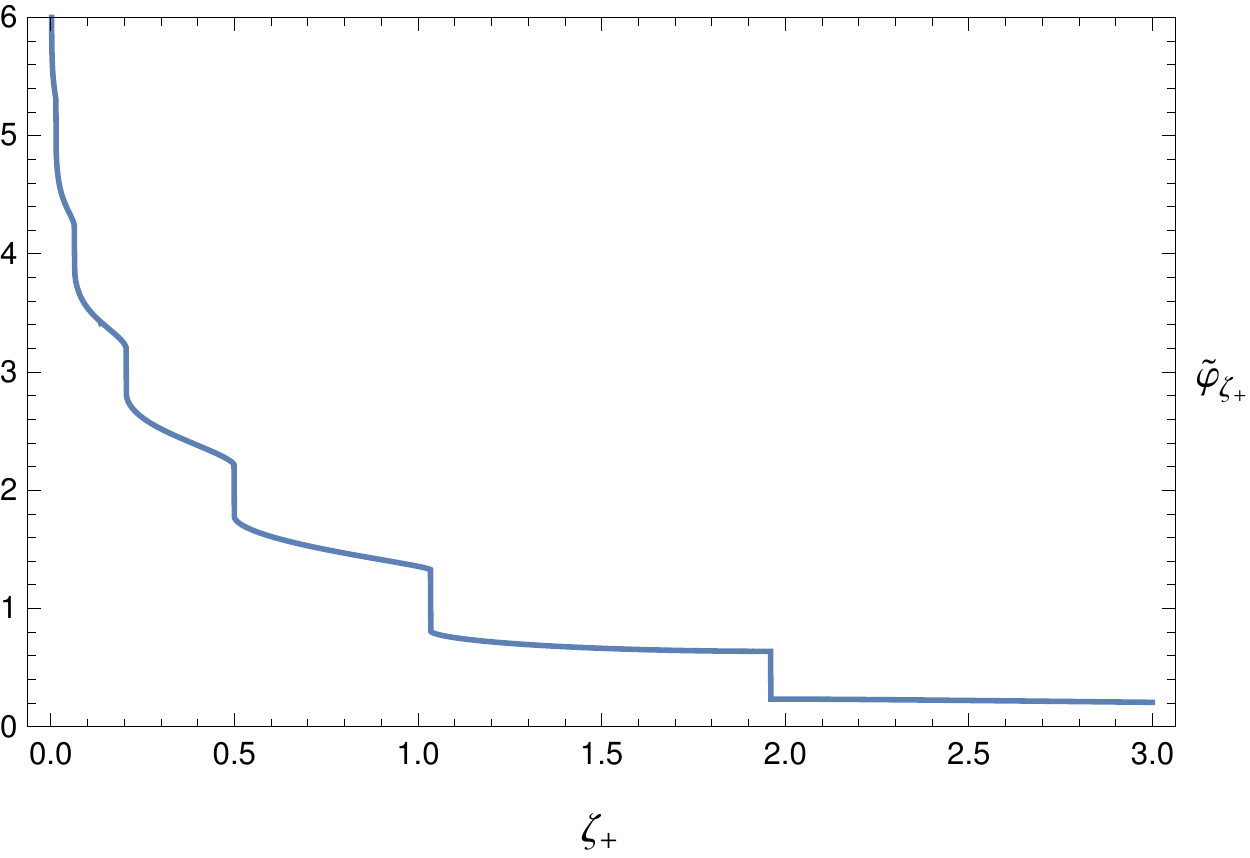}
 \caption{
 Location of the singularities of $\tilde{\varphi}_{\zeta_+}$ as a function of $\zeta_+$ by solving \eqref{susy_wp_ode} for $d=2$ and $\eta=1/10$ using the $n=1$ cutoff.
 The rightmost singularity corresponds to a Gaussian solution. The next singularity on the left is the tricritical Ising model. All other solutions with increasing \emph{odd} criticality are arranged
 from there towards the origin.
 }
 \label{susy_tri_spike2dplot}
\end{figure}

We now turn our attention to the boundary condition \eqref{wp_bc_even_w} 
corresponding to even solutions $w(\varphi)$ of the fixed point
equation. For this condition it is easy to see that 
$\eta_0$ in \eqref{susy_eta0} vanishes such that
LPA$^0$ becomes LPA. Furthermore, the integration 
of \eqref{susy_wp_ode} for $d=2$ and $\eta=0$ does not meet
any singularities in analogy to what happened in both Sect.s \ref{subsection_ising2d} and \ref{subsection_susy_ising2d}
because the canonical dimension of the field is zero. In Fig.\ \ref{susy_tri_spike2dplot} we plot the endpoint of the integration of
\eqref{susy_wp_ode} for $d=2$, boundary conditions \eqref{wp_bc_even_w} and trial anomalous dimension $\eta=1/10$
as a function of the parameter $\zeta_+$.
The plot is qualitatively similar to Fig.\ \ref{susy_spike2dplot},
displaying the same type of singularities in the form of discontinuities of $\tilde{\varphi}_{\zeta_+}$
across which the first derivative does not change sign.
A closer inspection of the solutions shows that the rightmost singularity 
seems to represent a transition between a free theory 
with quadratic potential for large values of $\zeta_+$ and a solution
with two maxima away from $\varphi=0$.

When moving towards the second singularity from the right two additional
minima of the potential appear
as expected for a (supersymmetric) tricritical Ising model.
Thus for $\zeta_+\downarrow \zeta_{+, {\rm cr}}$ we end up with
the supersymmetric extension of the tricritical Ising model.
The only available scheme is the LPA' for the cutoff $n=2$, for which we must tune the anomalous dimension and obtain
\begin{equation}
 \begin{split}
\zeta_+({\rm LPA}'{}_{n=2})=0.3793822032\,, &\quad \eta\phantom{'}=0.320147\,.
 \end{split}
\end{equation}
The values for all critical parameters are given in Tab.\ \ref{table_susy_triising2d},
while the critical exponents for even and odd deformations are listed in Tables \ref{table_susy_triising2d_even_exponents}
and \ref{table_susy_triising2d_odd_exponents}, respectively.
\begin{table}[h!]
\begin{tabular}{| l || r | r | r | r | r | r | r |}
 \hline
 & $w'(0)$ & $\sigma_{\rm cr}$ & $g_{\rm cr}$ & $g_{6,{\rm cr}}$ & $\eta_{\rm in}$ & $\eta_{\rm out}$ & $\nu$
 \\
 \hline
 LPA'${}_{n=2}$ & 0.3794 & 0.1439 & 9.16 & 15.36 & 0.3201 & 0.3201 & 1.6653
 \\
 \hline
\end{tabular}
\caption{Critical parameters for the supersymmetric tricritical Ising class in $d=2$.}
\label{table_susy_triising2d}
\end{table}
\begin{table}[h!]
\begin{tabular}{| l || r | r | r | r | r |}
 \hline
Shooting & $\theta^+_1$ & $\theta^+_2$ & $\theta^+_3$ & $\theta^+_4$ & $\theta^+_5$
 \\
 \hline
 LPA'${}_{n=2}$ & 0.6005 & -0.3129 & -1.5243 & -3.0449 & -4.8593
 \\
 \hline
 \hline
 SLAC: & $\theta^+_1$ & $\theta^+_2$ & $\theta^+_3$ & $\theta^+_4$ & $\theta^+_5$
 \\
 \hline
 LPA'${}_{n=2}$ & 0.6005 & -0.3129 & -1.5245 & -3.0465 & -4.8710
 \\
 \hline
\end{tabular}
\caption{Critical exponents of the even deformations for the on-shell potential of the supersymmetric tricritical Ising class in $d=2$.}
\label{table_susy_triising2d_even_exponents}
\end{table}
\begin{table}[h!]
\begin{tabular}{| l || r | r | r | r | r |}
 \hline
Shooting: & $\theta^-_1$ & $\theta^-_2$ & $\theta^-_3$ & $\theta^-_4$ & $\theta^-_5$
 \\
 \hline
 LPA'${}_{n=2}$ & 0.8399 & 0.1601 & -0.8780 & -2.2468 & -3.9170
 \\
 \hline
 \hline
 SLAC: & $\theta^-_1$ & $\theta^-_2$ & $\theta^-_3$ & $\theta^-_4$ & $\theta^-_5$
 \\
 \hline
 LPA'${}_{n=2}$ & 0.8399 & 0.1601 & -0.8780 & -2.2473 & -3.9213
 \\
 \hline
\end{tabular}
\caption{Critical exponents of the odd deformations for the on-shell potential of the supersymmetric tricritical Ising class in $d=2$.}
\label{table_susy_triising2d_odd_exponents}
\end{table}

As for the supersymmetric extension of the Ising class in Subsect.\ \ref{subsection_susy_ising2d}, two critical exponents
can already be predicted with the help of the results of Appendix \ref{appendix1}. 
In the present case the superpotential has the opposite
parity as for the Ising class and
both deformations discussed in the Appendix appear as odd deformations 
of both the potential and the superpotential.
The new relations in fact determine the first two exponents in Tab. \ref{table_susy_triising2d_odd_exponents} and take the values
\begin{equation}
 \theta^-_1 = \frac{2-\eta}{2}\,, \qquad \theta^-_2 = \frac{\eta}{2}\,,
\end{equation}
which hold true within our numerical precision.
The first relation is the equivalent of the superscaling relation \eqref{superscaling};
however, here it does not determine the first critical exponent of the even deformations
and thus does not give a direct relation between $\nu$ and $\eta$.

\section{Critical dimensionalities}

\label{section_critical_dimensionalities}

The two flow equations \eqref{vdot} and \eqref{susy_wdot} admit an analytic 
continuation to arbitrary fractional dimensions $d>2$,
leading to the question of whether it is possible to generalize the multicritical models to spaces with fractal properties.
This possibility has been investigated at length for the simple scalar model \eqref{vdot} in \cite{Codello:2012sc}, whose results we partly reproduce below for completeness.
However, the same is not true for the continuation of the WZ-model
with one supercharge.
One important physical motivation to continue the system to fractional dimensions is that
it was observed in \cite{Defenu:2014bea} that there is a correspondence between the scalar models in fractional dimensions with local (short-range) interactions,
and the same models in two or three dimensions with non-local (long-range) interactions.
It is thus an intriguing possibility that the analytic continuation of our Wess-Zumino model to fractional dimensions
plays a role in describing long-range effects in the phenomenon of emergent $2+1$ supersymmetry in topological superfluid phases \cite{Grover:2013rc}.
The interesting question to answer now is the following: What are the critical dimensions 
at which new critical models appear when continuously following $d$ from $4$ 
(where very probably there is only the Gaussian model) to $2$ (where there are infinitely many models)?

The analytic continuations of the WZ-model with
more than one supersymmetry
had already been studied in the literature in \cite{Giombi:2014xxa}
with the purpose of interpolating between the $a$- and $F$-theorem,
and in \cite{Bobev:2015jxa} where the conformal bootstrap of the model was considered.
The corresponding models all contain (on-shell) two real scalar 
fields, have two supersymmetries in $3$ dimensions
and relate to superconformal ${\cal N}=2$ models in $2$ dimensions.
This family of models is different from those considered in the present work.
 Our models have (on-shell) one scalar field and 
one supersymmetry in $3$ dimensions and relate to superconformal
${\cal N}=1$ models in $2$ dimensions.

Motivated by both the Landau-Ginsburg theory and the results 
of the local potential approximation, we assume that 
the local operators in the effective potential are the main actors 
when it comes to developing a new criticality class.
A new (upper) critical dimension appears every time a new local operator becomes marginal.
To show this, let us first concentrate on the scalar field model 
and equation \eqref{v_ode}.
The set of simple local operators that satisfy the boundary conditions is 
exhausted by the powers $\varphi^{2n}$
for $n \in \mathbb{N}$. Each of these local operators becomes marginal when the corresponding coupling
is canonically dimensionless; the dimensionality at which this happens is thus critical.
In the language of \eqref{v_ode} the most appropriate way to find the 
critical dimension is by observing that an operator is marginal if
\begin{equation}\label{marginal}
 \left.{\cal S}_{v,\eta}[v,v';\varphi]\right|_{v(\varphi)=\varphi^{2n}}=0\,,
\end{equation}
which can be explicitly solved for $d$ to obtain the critical dimensions
\begin{equation}\label{critical_dimensions}
 d_n = \frac{2n}{n-1}-\frac{n}{n-1}\eta\,.
\end{equation}
The series of critical dimensions accumulates at $d=2$ for $n\to\infty$ when $\eta=0$.

\begin{figure}[htpb]
 \vspace{0.5cm}
 \includegraphics[width=8cm]{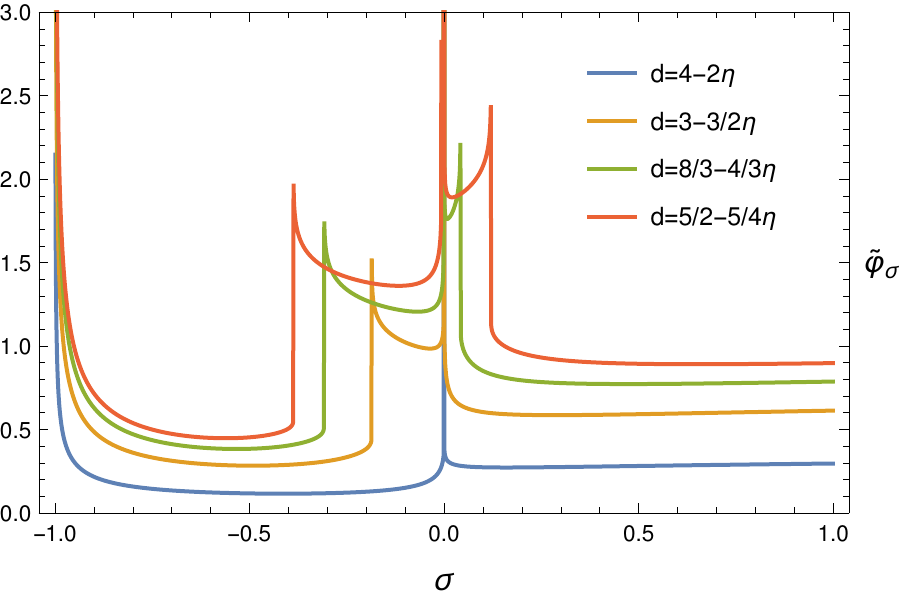}
 \caption{
 Plot of the function $\tilde{\varphi}_\sigma$ for the first few critical dimensions $d_n$.
 The plot was obtained using the value $\eta=1/10$.
 }
 \label{plot_critical_dimensions_scalar}
\end{figure}

The plots of the end points $\tilde{\varphi}_\sigma$ of the integration 
of \eqref{v_ode} for the first few critical dimensions are depicted
in Fig.\ \ref{plot_critical_dimensions_scalar}. Each curve can be deformed 
continuously into the others by changing $d$ in \eqref{v_ode}.
It is easy to see that, by lowering the dimension, 
new singularities appear in the plot corresponding to new
critical models. The way in which new singularities appear is as follows:
Let us start from the upper critical dimension $d_2=4-2\eta$ of the 
$\varphi^4$ model where a new singularity appears to the left of $\sigma=0$.
It corresponds to the Wilson-Fisher fixed point of the Ising class
which we already studied in Subsect.\ \ref{subsection_ising3d}.
For $d=d_2-\epsilon$ with infinitesimal $\epsilon$ this singularity is 
arbitrarily close to the Gaussian fixed point solution
that is generally considered within perturbation theory.
If we follow $d$ further down to $d_3=3-3/2\eta$ 
the singularity has moved left to a finite value of $\sigma$ and 
just below $d_3$  a new singularity appears to 
the right of $\sigma=0$ corresponding to the tricritical Ising model.
Again, this singularity can be followed for decreasing values of $d$ while it 
moves further to the right. The picture that we just described can be 
iterated further: Each time a new critical dimension is crossed a new 
singularity emanates from $\sigma=0$.
The singularities appear with an alternating sign of $\sigma=v''(0)$,
and this explains the ordering of the models that we encountered, for example, 
in Fig.\ \ref{spike2dplot}.
In each interval $d_{n+1}<d<d_n$ a new operator $\varphi^{2n}$ enters from its upper critical dimension defined through \eqref{marginal};
thus, it is able to satisfy the condition \eqref{singularity0}, which in turn implies a new branching of Fig.\ \ref{plot_critical_dimensions_scalar}
through the condition \eqref{singularity}.

Now we are ready to extend the above discussion to the supersymmetric
models. Recalling the form of the superpotential in \eqref{wz_effective_average_action},
a new critical dimension is expected to occur when a new operator $\Phi^{1+m}$ becomes marginal, for which
the symmetry of the model requires that $m\in \mathbb{N}$.
Using the notation introduced in \eqref{susy_wdot}, we thus seek the dimensionality for which
\begin{equation}\label{marginal_susy}
 \left.{\cal S}_{w,\eta}[w,w';\varphi]\right|_{w(\varphi)=\varphi^{m+1}}=0\,,
\end{equation}
which can be solved for $d$ to obtain the new critical dimensions
\begin{equation}\label{susy_critical_dimensions}
 d^{\,\rm susy}_m = \frac{2m}{m-1}-\frac{m+1}{m-1}\eta\,.
\end{equation}
Again, the critical dimensions define a series which accumulate at $d=2$ 
for $m\to\infty$ when $\eta=0$.
The two sets of critical dimensions \eqref{critical_dimensions} and \eqref{susy_critical_dimensions} strictly
coincide for $\eta=0$, while the difference is related to the fact that the on-shell superpotential defined in \eqref{wz_effective_average_action_on_shell}
is subject to a multiplicative wavefunction renormalization which has no analogue in 
the scalar field theory.

\begin{figure}[htpb]
 \vspace{0.5cm}
 \includegraphics[width=8cm]{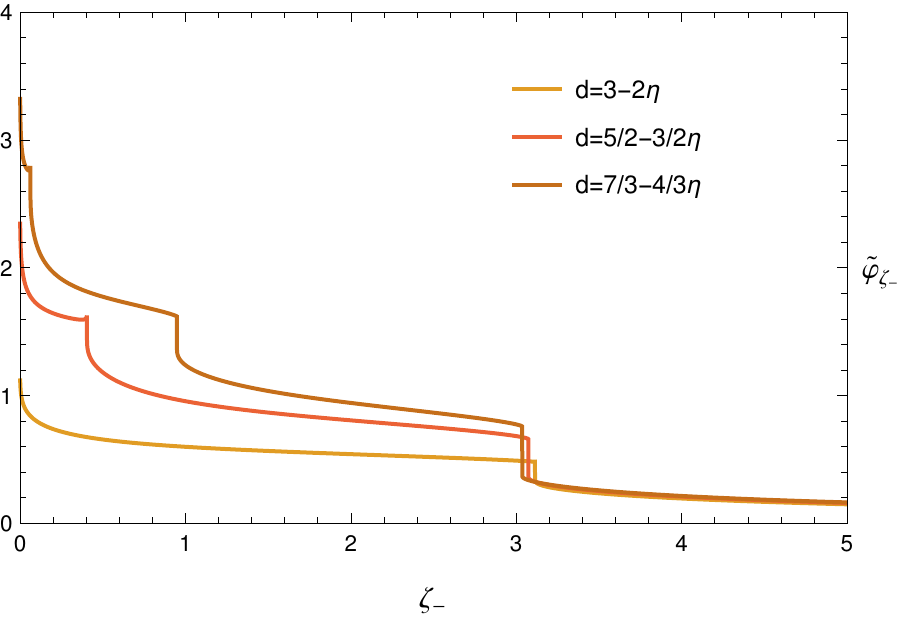}
 \caption{
 Plot of the function $\tilde{\varphi}_{\zeta_-}$ for the first few critical dimensions $d^{\,\rm susy}_m$.
 The plot was obtained using the value $\eta=1/10$.
 }
 \label{plot_critical_dimensions_susy1}
\end{figure}

\begin{figure}[htpb]
 \vspace{0.5cm}
 \includegraphics[width=8cm]{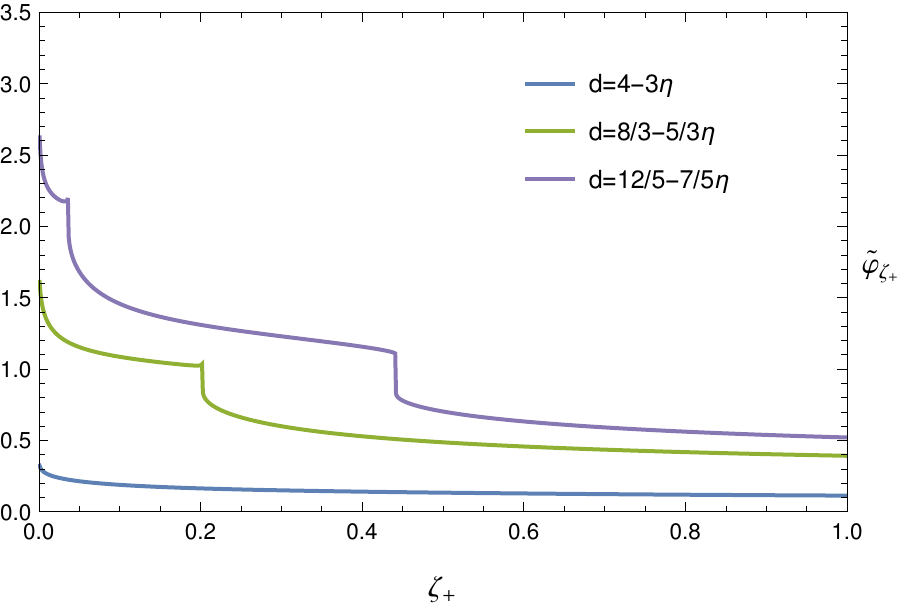}
 \caption{
 Plot of the function $\tilde{\varphi}_{\zeta_+}$ for the first few critical dimensions $d^{\,\rm susy}_m$.
 The plot was obtained using the value $\eta=1/10$.
 }
 \label{plot_critical_dimensions_susy2}
\end{figure}

Now we can follow the solution of \eqref{susy_wp_ode} with boundary conditions \eqref{wp_bc_even_w} or \eqref{wp_bc_odd_w} as a function of the dimension
when lowering $d$ from $d^{\,\rm susy}_2=4-2\eta$ to $d^{\,\rm susy}_\infty=2-\eta$. The result is depicted in Fig.s \ref{plot_critical_dimensions_susy1} and \ref{plot_critical_dimensions_susy2}
which show the function $\tilde{\varphi}$ for even and odd boundary conditions 
for the superpotential.
We see that the dimensions $d^{\,\rm susy}_m$ for $m$ \emph{odd} are critical 
for the plot in Fig.\ \ref{plot_critical_dimensions_susy1}, while
those for $m$ \emph{even} are critical for the plot in Fig.\ \ref{plot_critical_dimensions_susy2}. In both cases, by lowering $d$
a new singularity is produced in the plot when a new critical dimension is crossed.
The alternation of the singularities between the Fig.s \ref{plot_critical_dimensions_susy1} and \ref{plot_critical_dimensions_susy2}
is analogous to the one that occurs between the left and right sides of Fig.\ \ref{plot_critical_dimensions_scalar}
and was expected on the basis of the relation \eqref{susy_sigma} between the parameters $\sigma$ and $\zeta_\pm$.
Since in our approximation only the operators of the form
$\varphi^{n}(\bar\psi\psi)^m$ can determine the occurrence of new critical dimensions,
in the limit of zero anomalous dimension our result \eqref{susy_critical_dimensions}
coincides with the set of critical dimensions given in \cite{Vacca:2015nta}
where a simple Yukawa model is considered. For this reason, it is tempting to speculate
that the Yukawa and supersymmetric universality classes are the same.
This, however, is true only to some extent,
as the allowed deformations of the Wess-Zumino model
are constrained by supersymmetry if compared to those of a Yukawa system,
therefore changing the spectrum of the theory
(see \cite{Bashkirov:2013vya} for comments on this model from the point of view of the conformal bootstrap).

\section{Discussion}

\label{section_discussion}

We studied the functional renormalization group flow of simple scalar 
field theories and their ${\cal N}=1$ supersymmetric extensions in various local potential approximation schemes
using a scaling solutions approach,
pioneered in \cite{Morris:1994ie,Morris:1996kn,Morris:1994ki}, in order 
to determine the stationary solutions of the flow.
Our approach to the scalar model extends the work in \cite{Codello:2012sc}, while the results for the supersymmetric Wess-Zumino model are new.
Both families of models exhibit the expected universality 
classes in terms of stationary solutions of the flow.
In three dimensions we confirm the presence of both the Ising class and its supersymmetric counterpart,
while in two dimensions infinitely many stationary solutions can be observed
which we identify with the minimal models of CFT and super-CFT.
It is rather exceptional that the infinite tower of minimal models is visible within
such a simple truncation of the renormalization group flow that contains only a local potential
and an anomalous dimension for the field.

At various stages, we offered a new understanding of the mechanism 
with which new critical solutions
appear as stationary solutions of the flow of the local potential.
The key idea is tightly connected to (and best visualized through) the 
Ginzburg-Landau theory:
New critical models appear whenever new local operators turn into relevant deformations
of the spectrum of the theory. We explicitly observe that the appearance of each model
is governed by the canonical scaling of a specific operator which can be pinpointed, for example,
by analytically continuing the renormalization group flow between two and three dimensions.
The critical models can be identified as discontinuities in the plot of the domain of existence
of all possible models using both scaling arguments and the Landau description
that counts the number of nodes of the solutions \cite{ZinnJustin:2002ru}.

For the sake of the exposition,
we concentrated our attention on the Ising and tricritical Ising classes, and their ${\cal N}=1$ supersymmetric counterparts.
Most importantly, we introduced two new methods to compute the relevant deformations of these models:
The shooting and the pseudo-spectral SLAC methods.
While the shooting method has already been applied in \cite{Codello:2014yfa} to compute only the critical exponent
of the correlation length, we showed that it can be used quite effectively 
to obtain many low-lying exponents.
One important ingredient of the present work is the application 
of the SLAC derivative method
to the computation of the spectrum. This latter approach can in fact 
easily produce a large number of critical exponents
with one single calculation, effectively beating all other computationally slower algorithms
and showing a mild dependence on the boundary conditions.
The only limitation of the SLAC derivative method is in the size of the numerically determined critical solution,
which is finite in our implementation. We thus expect the method to work even better when global solutions
are available, such as in \cite{Borchardt:2015rxa}.

The functional renormalization group method is an efficient tool to 
analytically continue the flow in parameters that are
fixed in most alternative approaches, for example lattice simulations,
such as the dimensionality of the system.
We thus provided an analytic continuation of the ${\cal N}=1$ Wess-Zumino model 
between two and three dimensions that, to the best of our knowledge, is new.
This continuation can continuously follow the stationary solutions of the 
local potential flow from three to two dimensions, showing explicitly 
what the critical dimensions are at which new models appear.
Without wave function renormalization the critical dimensions of the 
supersymmetric Wess-Zumino model agree with those of the Yukawa model 
with the same number of on-shell degrees of 
freedom calculated in \cite{Vacca:2015nta}. It is an interesting 
open question whether this remains true beyond the LPA truncation
and, in particular, with non-zero anomalous dimensions.

A particular feature of our work is that we can give a self-consistent
description of any critical model within a very simple local potential approximation of the renormalization group flow.
We provided an extensive and detailed numerical analysis of Ising and tricritical Ising models.
Our numerical results are in good agreement with 
those obtained by dedicated Monte Carlo simulations and by CFT
for the critical exponent of the correlation length $\nu$,
but typically fail in determining the anomalous dimension $\eta$ accurately.
A better determination of the anomalous dimension within
the functional approach is crucial for improving the accuracy of the method.
It is natural to compare our results for the simple scalar field
theory with Morris' contributions to the derivative expansion of
the renormalization group \cite{Morris:1994ie,Morris:1994ki},
in which the local potential approximation is complemented by a field-dependent wavefunction renormalization
and for which an additional scaling symmetry (due to the choice of a powerlaw cutoff)
allows to determine the anomalous dimension from the boundary conditions of the flow.
We find a qualitative agreement with the results of \cite{Morris:1994ie,Morris:1994ki}
for both the Ising and tricritical Ising classes. As expected,
the agreement becomes increasingly worse when the field dependent 
wavefunction renormalization
of \cite{Morris:1994ie,Morris:1994ki} deviates from the value one.
The lesson to be learned here is that, while the local potential approximation works
extraordinarily well in showing that the critical models are visible in our functional renormalization
group approach, better numerical estimates of the critical parameter can only be achieved
by enlarging the local potential truncation to include further operators.
Therefore, an extension of our analysis to truncations of the effective action which account
for a larger space of operators in functional form, such as a field dependent wavefunction renormalization,
represents the most important and compelling prospect of our work.

Another natural comparison for our work would be the recent developments
on the conformal bootstrap approach \cite{ElShowk:2012ht,El-Showk:2013nia}.
The latter uses Ward-identities of conformal field theory
to constrain the operator algebra and provide
accurate predictions for the critical exponents of the Ising model.
In our approach, however, conformal invariance is an outcome of the requirement 
of scale invariance
by spotting stationary solutions of the renormalization group flow and applying the Landau classification to the solutions.
Thus it is tempting to speculate that in a functional renormalization group approach 
it might not be necessary to use the full power of conformal field theory, 
but simply enhance the correlators appropriately
to correctly capture the long range correlations in the vicinity of a critical point.
Such an extension of the functional truncation might roughly follow the lines of
\cite{Codello:2015lba},
or be dictated by the desire of maintaining two-loops universality when continuing the
results to four dimensions \cite{Codello:2013bra} (see also \cite{Rychkov:2015naa} for a conformal field theory perspective).

\paragraph*{Acknowledgments.}

We are grateful to T.\ R.\ Morris and L.\ Zambelli for valuable comments on an earlier draft of the paper.
The research was supported by the Deutsche Forschungsgemeinschaft (DFG) graduate school GRK 1523/2.
A.\ Wipf thanks the DFG for supporting this work under grant no.\ Wi 777/11-1.
O.\ Zanusso thanks the DFG for supporting this work under grants no.\ Gi 328/7-1 and Gi 328/6-2 (FOR 723).

\begin{appendix}

\section{Magnetic and derivative fluctuations}

\label{appendix1}

Let us consider a general renormalization group evolution equation for a scale dependent ``potential'' $h(\varphi)$ of the form
\begin{equation}
 \begin{split}\label{hdot}
  &k\partial_k h(\varphi)
  =
  {\cal S}_{h}[h,h';\varphi]+{\cal F}_{h}[h'']\,,
  \\
  &{\cal S}_{h}[h,h';\varphi]\equiv-d_h h(\varphi)+ d_\varphi\, \varphi\, h'(\varphi)\,.
 \end{split}
\end{equation}
As done for \eqref{vdot} we introduced a scaling part ${\cal S}_{h}[h,h';\varphi]$
that contains some general (possibly almost canonical) dimensions $d_\varphi$ and $d_h$ for the dimensionful counterparts of the field $\varphi$ and the potential itself respectively.
We assume that the non-trivial part of the flow ${\cal F}_{h}[h'']$ only depends on $h''(\varphi)$ and is otherwise arbitrary.
Let us also introduce a $k$-stationary solution $h^*(\varphi)$ of \eqref{hdot} of given parity.
The linearized equation for the fluctuations $\delta h(\varphi)$ defined by
\begin{equation} 
 h(\varphi)\to h^*(\varphi)+\epsilon\,\delta h(\varphi)\,\left(\frac{k}{k_0}\right)^{-\theta}
\end{equation}
takes the form
\begin{equation}\label{linearized_hdot}
 \left(\theta-d_h\right)\delta h(\varphi)+d_\varphi\varphi \delta 
h'(\varphi)+{\cal F}^\prime_{h}[h^{*\prime\prime}]\delta h''(\varphi)=0\,,
\end{equation}
in which the prime denotes the derivative with respect to the argument.

For any $h^*(\varphi)$ with fixed parity there always exist
two fluctuation modes with fixed parity and critical exponents that are
directly related to $d_\varphi$ and $d_h$.
The first one is the magnetic deformation $\delta h(\varphi)\propto \varphi$.
It is easy to see by direct substitution that this deformation 
is normalized at the origin by $\delta h'(0) = 1$ and
solves \eqref{linearized_hdot} for the critical exponent $\theta = (d_h-d_\varphi)$.

The other fluctuation is $\delta h(\varphi)\propto h^{*\prime}(\varphi)$. 
To see this and to determine the corresponding 
critical exponent one computes
\begin{equation}
 \begin{split} 
  \frac{\rm d}{{\rm d}\varphi}\Bigl({\cal S}_{h}[h^*,h^{*\prime};\varphi]+{\cal F}_{h}[h^{*\prime\prime}]\Bigr)=0\,,
 \end{split}
\end{equation}
and compares the resulting expression with the fluctuation equation \eqref{linearized_hdot}.
The comparison reveals that the critical exponent of this second deformation is
$\theta = d_\varphi$. The deformation has opposite parity compared to
the fixed point solution $h^*(\varphi)$.

In the specific example of the scalar field theory \eqref{vdot} these 
two deformations have critical exponents
\begin{equation}
 \frac{d+2-\eta}{2}\,\,\,\,{\rm and}\,\,\,\, \frac{d-2+\eta}{2}\,.
\end{equation}
The ratio of the second over the first exponent is the well known 
thermodynamical exponent $\delta$. The exponents always show up in our 
numerical computations. Since in the scalar case the fixed point solution
$h^*=v^*$ is even, both exponents are contained in the spectrum
corresponding to the odd fluctuations.
Given the simple relation of these exponents with the anomalous dimension 
they can be used, in principle, to test the quality of the numerical algorithm 
used when studying the fluctuation spectrum.

\section{The LPA' matching algorithm}

\label{appendix_matching_algorithm}

We describe here the matching algorithm that was applied for obtaining the LPA' solutions in greater detail. Similarly as in the
Introduction, we do this for the three dimensional Ising class, but the considerations
will apply straightforwardly to all other examples.

\begin{figure}[htpb]
 \vspace{0.5cm}
 \includegraphics[width=7cm]{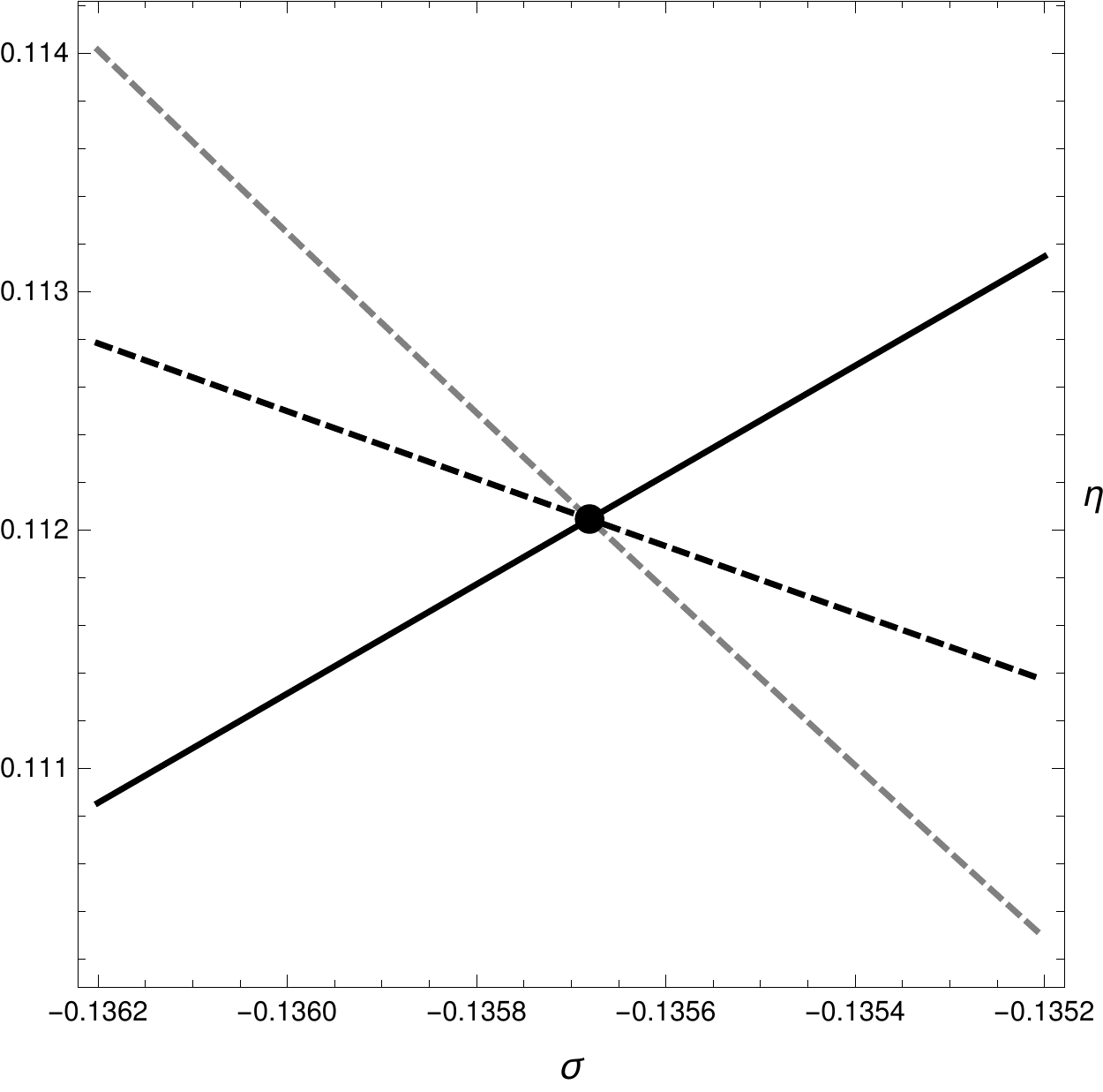}
 \caption{
 Determination of $\eta$ in the LPA' of the 3$d$ Ising model through the 
 matching condition $\eta_{\rm in} = \eta_{\rm out}$.
 The solid curve parametrizes the location of $\sigma=\sigma_{\rm cr}$ 
 as a function of $\eta_{\rm in}$,
 while the dashed curve is a plot of $\eta_{\rm out}$ of the solution 
 corresponding to $\sigma_{\rm cr}$ for the given $\eta_{\rm in}$.
 The gray dashed curve represents the anomalous dimension of solutions 
 for which $\eta_{\rm in} = \eta_{\rm out}$, but it is
 computed by integrating non critical values of $\sigma$.
 The dot gives, by construction, a solution which is both critical and 
 satisfies the matching condition.}
 \label{matching3dplot}
\end{figure}

As discussed in the main text, the stationary solutions of equation \eqref{vdot} depend parametrically on the anomalous dimension and the required input value 
when solving this equation was denoted by $\eta_{\rm in}$.
For each value of the parameter $\eta_{\rm in}$ we can compute a corresponding value of $\sigma_{\rm cr}$
by localizing the appropriate singularity in the plot of the value $\tilde{\varphi}_\sigma$.
For reasonably small values of $0\leq \eta_{\rm in} \lesssim 1 $ the plot of $\tilde{\varphi}_\sigma$
is qualitatively similar to Fig.\ \ref{spike3dplot}. In Fig.\ \ref{matching3dplot} we plot as a solid line
the curve parametrizing the location of $\sigma_{\rm cr}$ as a function of $\eta_{\rm in}$.
For our definition of LPA' scheme, we need the anomalous dimension computed at the minimum $\varphi_0$ of the solution,
which we dubbed $\eta_{\rm out}$, to coincide with $\eta_{\rm in}$.
This is not always the case, as shown in Fig.\ \ref{matching3dplot} where the value of $\eta_{\rm out}$ is plotted
as a dashed curve.
The mapping $\sigma\to\eta_\mathrm{out}$ is only defined locally, as its
definition uses information about the potential at its (absolute) minimum,
which is not always reached by the numerical solution. 
The intersection of the two curves gives, by definition, the condition 
$\eta_{\rm in}=\eta_{\rm out}$, and thus a critical solution in the sense of the LPA'.

The scales of Fig.\ \ref{matching3dplot} are particularly instructive, as they
make clear that any numerical error in $\sigma_{\rm cr }$ induces
an error of $\eta$ which is at least one order of magnitude bigger. Among 
the systems studied in this paper, the $3d$ Ising model is the one that shows 
the smallest  error propagation from $\sigma_{\rm cr}$ to $\eta$. In other cases
the amplification is by two or three order of magnitudes.
Since $\sigma_{\rm cr}$ is the most important parameter
as it controls the precision with which all other parameters are determined, 
we decided to determine it with precision of about $10^{-10}$ or higher. 
The anomalous dimension $\eta$ is then typically determined at a
precision in the range $10^{-7}-10^{-9}$ depending on the model considered.

The Newton-type algorithm for the matching works as follows. Once a 
sufficiently accurate trial value for $\eta_{\rm in}\simeq \eta$(LPA') is found (for example with precision $10^{-3}$ as seen in Fig.\ \ref{matching3dplot}),
this value is used to determine $\sigma_{\rm cr}$ and subsequently $\eta_{\rm out}$. If $\eta_{\rm in}$ and $\eta_{\rm out}$
do not match to the desired precision, the value of $\eta_{\rm in}$ 
is changed in the direction of $\eta_{\rm out}$
by a step smaller than the desired precision
and the procedure is then iterated. The step size should be smaller 
than the desired precision; otherwise, occasionally the algorithm might get stuck in loops,
which are caused by the fact that the two curves of Fig.\ \ref{matching3dplot} are almost parallel if plotted with the same scale on both axes.
This very simple algorithm yields precise estimates for $\sigma_{\rm cr}$.

\section{Continuity of the critical exponents}

\label{appendix_sigma_dependence}

When one tries to calculate the fluctuation spectrum of a given
fixed point solution one meets the problem that the solution 
of the fixed point equation is only known up to a maximal value 
$\tilde{\varphi}_{\sigma}$ of the 
field (cf.\ the corresponding discussion in Subsect.\ \ref{subsection_scaling_solutions}).
This is due to discretization and rounding errors involved in the 
numerical computation of the 
critical values of the initial 
conditions.\footnote{Spectral methods do better in finding
global solutions \cite{Heilmann:2014iga,Borchardt:2015rxa}.}
Even for initial conditions extremely close 
to criticality one finds $\tilde{\varphi}<\infty$.

\begin{figure}[htpb]
%  \vspace{0.5cm}
 \includegraphics[width=8cm]{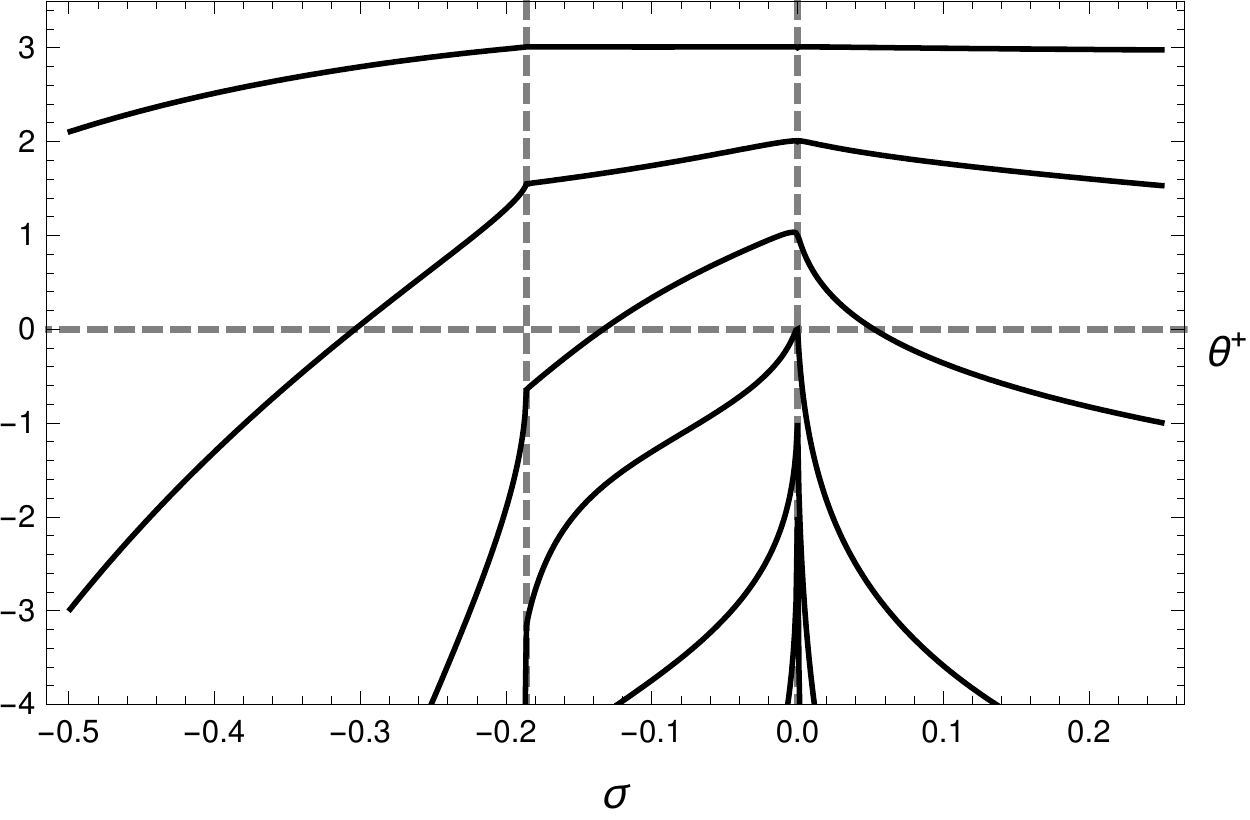}
 \caption{
 Dependence on $\sigma$ of the critical exponents $\theta^+_i$ for $i=0,\dots,6$ from top to bottom
 of the three dimensional Ising class.
 This dependence is computed using the shooting algorithm.
 }
 \label{sigma_dependence_plot}
\end{figure}

Both the shooting and spectral-SLAC methods have been 
adapted for (various) boundary conditions at the maximal value 
of the field $\tilde\varphi_\sigma$ for 
stationary solutions of equation \eqref{vdot} parametrized by $\sigma$
to extract the eigenfunction and eigenvalues of the corresponding 
fluctuation operator. This can be done for non-critical solutions
and the critical solution.
In Fig.\ \ref{sigma_dependence_plot} we plot the exponents 
corresponding to the parity even deformations
of the solutions of \eqref{v_ode} for $d=3$ with boundary 
conditions \eqref{v_zero}. The plot includes the exponent $\theta^+_0$
of the constant deformation that corresponds to the scaling of the volume 
operator. It was never listed in the tables since for all critical models 
it is just the dimension of the system.

The plot for the Ising class in Fig.\ \ref{sigma_dependence_plot} reveals 
many interesting features.
All exponents $\theta_i^+$ seem to be continuous functions of the 
parameter $\sigma$ with cusps at the critical values $\sigma=\sigma_{\rm cr}$ 
corresponding to the non-trivial fixed point and at
$\sigma=0$ corresponding to the Gaussian fixed point. The critical
values of $\sigma$ are marked with vertical dashed lines in the plot.
The critical exponents of all deformations are negatively divergent at $\sigma=-1$
and gradually increase with increasing curvature $\sigma$ to their value at the Ising class.
Beyond $\sigma_{\rm cr}$ they continuously interpolate
with the spectrum of the free theory at the Gaussian point, where they
attain their maximal values, and then their values decrease again.
Each eigenvalue shows a distinct behavior near the critical points.
Also note that the scaling exponent of the volume operator $\theta^+_0$
is not equal to $3$ below $\sigma_{\rm cr}$, but takes this value
at $\sigma_{\rm cr}$ and stays there up to $\sigma=0$.
Beyond this point it decreases again albeit very slowly.
All higher exponents approach $\sigma_{\rm cr}$ with a slope that differs
between the left side and the right side.
as is expected from the discontinuous change in the domain
of existence of the potential (see also Fig.\ \ref{discontinuity} which shows almost critical solutions from the left and the right of the critical point).

\section{The SLAC derivative method}

\label{appendix_slac}

We describe here the spectral-SLAC method for the computation of the critical exponents that was applied in the present work. Again we use the scalar 
field theory of Sect.\ \ref{section_simple_scalar} as an example, but all 
considerations can be extended to the supersymmetric model of 
Sect.\ \ref{section_wz_model}.
The stability analysis around the solution of the second order differential 
equation \eqref{v_ode} yields the linear second order differential equation
\eqref{linearized_vdot} that depends parametrically on the critical exponents $\theta$.
The critical exponents $\theta$ must thus, in general, be fixed by the boundary conditions.
If a global solution to \eqref{v_ode} is present, the physically meaningful condition needed is that the fluctuations are polynomially bounded in the limit of large field
\cite{Morris:1994ki}, which in turn implies a quantized spectrum of critical 
exponents with an upper bound $\theta\in\{\theta_i\}$.

Our numerical solutions, however, do not extend to the infinite field 
(see for example Fig.\ \ref{criticalsol}); therefore, we have
to impose boundary conditions at the extremum $\tilde{\varphi}$ of the domain 
of validity, rather than at infinity.
The influence of these auxiliary boundary conditions can be minimized by 
assuming periodicity, in which case the amplitude of the fluctuations 
can vary freely near $\tilde{\varphi}$ rather than being zero as enforced by the shooting method.
With these premises, there is a very fast, precise and efficient algorithm 
to compute the spectrum of the critical exponents $\theta_i$
which is known the spectral-SLAC method, because it uses the lattice's SLAC 
derivative \cite{Drell:1976bq}.

Let the linear ODE describing some fluctuations $\delta u(\varphi)$ around the critical solution $v^*(\varphi)$ be of the form
\begin{equation}\label{general_fluctuations_ode}
 \left(-a(\varphi) \partial_\varphi^2 +b(\varphi)\right) \delta u(\varphi) = -\theta \, \delta u(\varphi),\quad\vert\varphi\vert\leq\tilde\varphi\,,
\end{equation}
where we used the fact that any linear second order ODE
can be cast in a form without first order derivative 
(for the scalar model this transformation 
is given in \eqref{transformed_fluctuation}).
The functions $a(\varphi)$ and $b(\varphi)$ generally depend on $v^*(\varphi)$:\ Let us assume that the function $a(\varphi)$ is positive.

Eqn.\ \eqref{general_fluctuations_ode} can be interpreted as the quantum 
mechanical problem of diagonalizing the Hamiltonian ${\cal H}\equiv -a \partial_\varphi^2 +b$ with eigenvalues $-\theta$.
Now we discretize the interval $[-\tilde\varphi,\tilde\varphi]$ on 
which $\varphi$ is defined by $N$ lattice points
$\{\varphi_1,\dots,\varphi_N\}\equiv\Lambda$ with constant separation $2\tilde\varphi/N$.
The momentum takes its values in the dual lattice 
\begin{equation}
\Lambda^*=\left\{p_\ell=\frac{\pi}{\tilde\varphi}\left(\ell-\frac{1+N}{2}\right)\big\vert
\ell=1,2,\dots,N\right\}\,.
\end{equation}
The lattice's SLAC derivative is defined as
\begin{equation}
 \sum_{k'=1}^N \left(\partial^{\rm slac}_\varphi\right)_{kk'} f(\varphi_{k'})  =\frac{{\rm i}}{\sqrt{N}} \sum_{\ell=1}^N  p_\ell \,{\rm e}^{{\rm i}p_\ell \varphi_k} 
 \tilde{f}(p_\ell)\,,
\end{equation}
where $f$ is a complex-valued function on $\Lambda$ and
$\tilde{f}(p)$ is its Fourier transform. Differently from many standard lattice's derivatives, 
the non-local SLAC derivative is given by a matrix whose entries are,
in general, non-zero for any pair of lattice-points
$\varphi_k$ and $\varphi_{k'}$.

With respect to periodic boundary conditions the second derivative based
on the SLAC-derivative reads
\begin{equation}\label{slac_matrix}
\begin{split}
-\left(\partial^{\rm slac}_{\varphi}\right)^2_{kk}&=\left(\frac{\pi}{2\tilde\varphi}\right)^2\frac{N^2-1}{3}\\
-\left(\partial^{\rm slac}_{\varphi}\right)^2_{k\neq k'}&=
2\left(\frac{\pi}{2\tilde\varphi}\right)^2 (-1)^{k-k'}
     \frac{\cot\frac{\pi}{N}(k-k')}{\sin\frac{\pi}{N}(k-k')}
%   h=\partial_{w'}{\cal S}_{w',\eta}\,,
\end{split}
\end{equation}
with an odd number $N$ of lattice points. For anti-periodic boundary conditions 
(suitable for the parity-odd fluctuations) one uses the same matrix
but an even number $N$ of lattice points.
This matrix \eqref{slac_matrix} is used to 
discretize \eqref{general_fluctuations_ode} in the form
\begin{equation}\label{general_fluctuations_ode_discretized}
 \sum {\cal H}_{kk'}\, \delta u_{\varphi_{k'}} = -\theta \, \delta u_{\varphi_k} \,.
\end{equation}
The discretized operator ${\cal H}_{kk'}$ is an $N\times N$ matrix whose 
eigenvalues can be computed using standard
numerical methods. Up to a sign these are the exponents $\theta_i$.
The method yields very accurate eigenvalues if the corresponding
eigenfunctions have their main support within the 
domain $[-\tilde\varphi,\tilde\varphi]$ and if the
number of lattice points $N$ is sufficiently large to resolve the 
typical variations of the eigenfunction.
The biggest advantage of the SLAC method is the efficient
suppression of discretization
errors and (for sufficiently large $\tilde\varphi$ and $N$)
the very accurate results for the low lying $\approx N/3$ eigenvalues of the spectral
problem with just one diagonalization.
The accuracy can be estimated by 
varying both the number of lattice points 
and the discretization interval $[-\tilde\varphi,\tilde\varphi]$.

\section{The extended supersymmetric solution}

\label{appendix_crossing}

The critical supersymmetric solutions obtained in Subsections \ref{subsection_susy_ising3d}, \ref{subsection_susy_ising2d} and \ref{subsection_susy_tri_ising2d}
have a limited domain of existence, but can be extended to global ones by integrating through the point $\tilde{\varphi}$ defined by ${\cal F}^\prime_{w,\eta}=0$ in \eqref{susy_singularity2} and \eqref{susy_phimax}. When approaching 
the critical point the numerator of \eqref{susy_wp_ode}, that 
represents the scaling of $w'$, fulfills
\begin{equation}
 \frac{\rm d}{{\rm d}\varphi}{\cal S}_{w,\eta} = {\cal S}_{w',\eta} \to 0\, \quad {\rm when}\quad \varphi\to \tilde{\varphi}\,,
\end{equation}
thus balancing the corresponding zero of the denominator and giving rise to 
a finite $w'''(\varphi)$ in the limit.

\begin{figure}[htpb]
 \includegraphics[width=8cm]{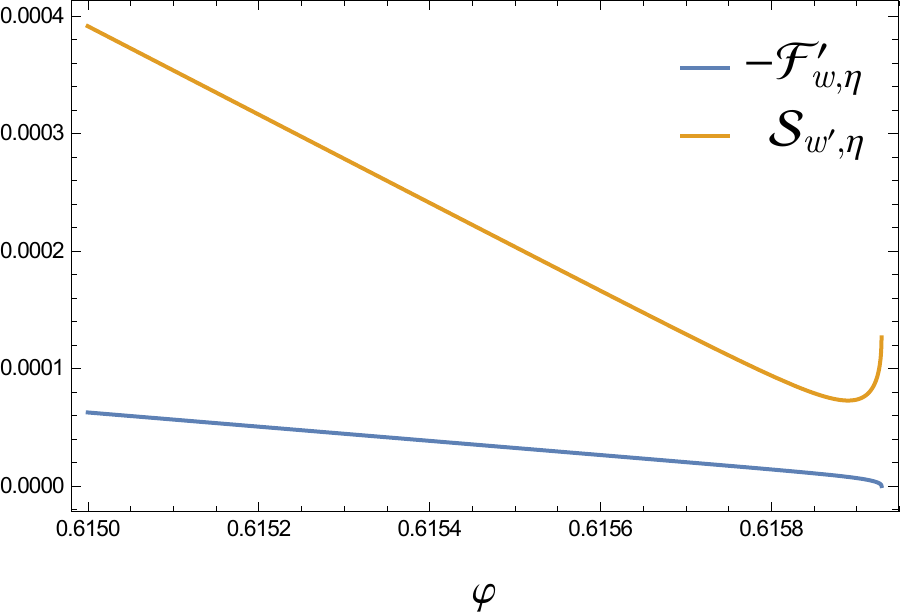}
 \caption{
The behavior of the functions ${\cal F}'_{w,\eta}$ and ${\cal S}_{w',\eta}$ near 
the removable singularity.
On purpose we chose a value of $\zeta_{-}$ away from criticality,
so that the numerical discrepancy between the two functions is easily seen.
 }
 \label{SprimeandFprime}
\end{figure}

Our strategy for numerically integrating beyond $\tilde{\varphi}$ is to 
perform a linear Taylor expansion of $w'(\varphi)$ and $w''(\varphi)$
for which ${\cal S}_{w',\eta}$ and ${\cal F'}_{w,\eta}$
change sign \emph{simultaneously} at the critical point.
This can be achieved by choosing the almost critical initial 
condition $\zeta_\pm \gtrsim \zeta_{\pm, {\rm cr}}$,
for which the solution will terminate when ${\cal F}^\prime_{w,\eta}\approx0$,
while ${\cal S}_{w',\eta}$ is still finite.
We illustrate in Fig.\ \ref{SprimeandFprime} the behavior of the 
functions ${\cal F}^\prime_{w,\eta}$ and ${\cal S}_{w',\eta}$ in the vicinity of $\tilde{\varphi}$
for the two dimensional supersymmetric Ising class found in Subsect.\ \ref{subsection_susy_ising2d} within the LPA' scheme.
The plot shows that the two functions almost reach zero at the same point, 
before deviating because of the numerical uncertainty in the determination of
the critical value of $\zeta_-$. The necessary condition for the existence of a 
finite and non-zero $w'''(\tilde{\varphi})$ is that the two functions intersect 
the $\varphi$-axis at the same point $\tilde\varphi$. Thus,
the crucial point is to accurately identify a small intervall
in Fig.\ \ref{SprimeandFprime} where ${\cal F}'_{w,\eta}$ and ${\cal S}_{w',\eta}$ 
are almost zero but still depend linearly on $\varphi$.
Then we perform a linear interpolation that ensures that both 
functions change sign and check that this happens at the same
point $\tilde\varphi$.

\begin{figure}[htpb]
%  \vspace{0.5cm}
 \includegraphics[width=8cm]{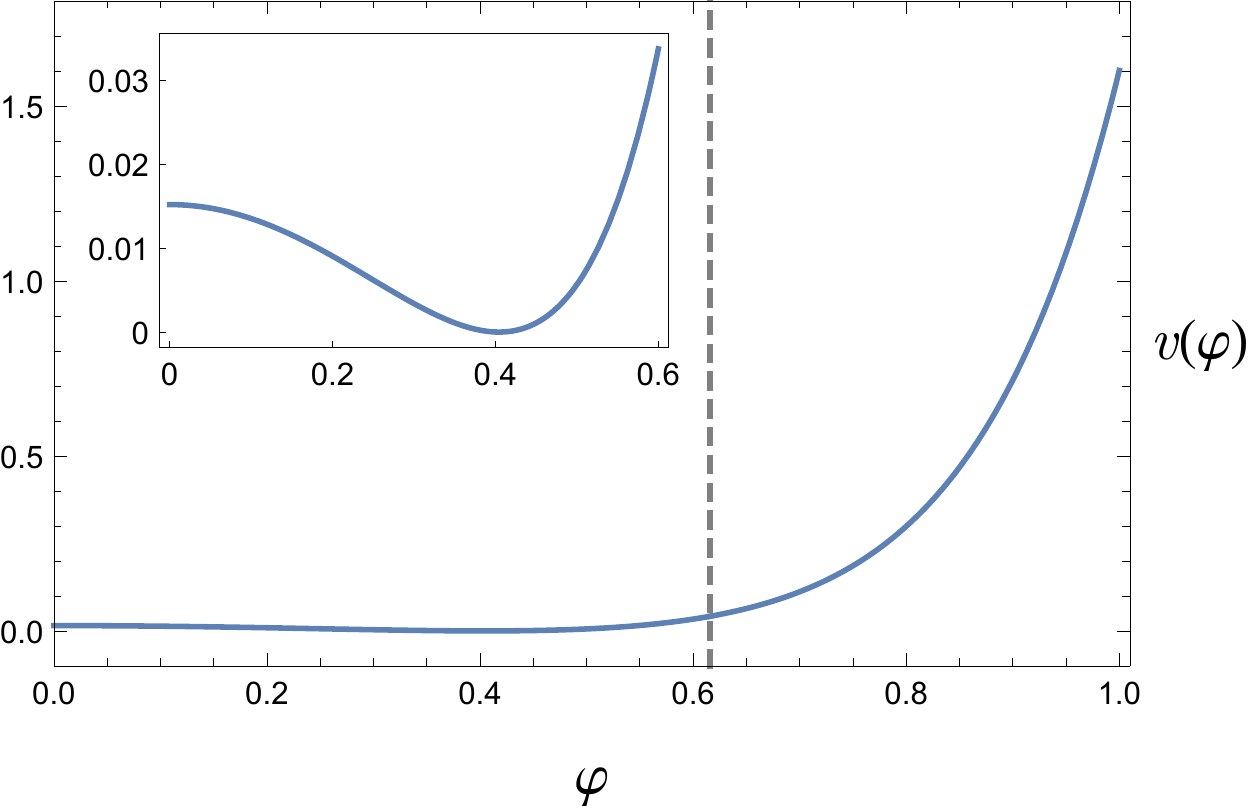}
 \caption{
 The global on-shell effective potential $v(\varphi)=w'(\varphi)^2/2$ for the LPA' solution of Subsect.\ \ref{subsection_susy_ising2d}.
 The vertical dashed line denotes the position of the singularity $\tilde{\varphi}$.
 }
 \label{globalsol2dsusy}
\end{figure}

Once the linear interpolation beyond $\tilde{\varphi}$ is made,
the numerical integration of \eqref{susy_wp_ode} yields
the solution for all values of the field. For large fields
the numerator and denominator slowly approach zero thus realizing 
the predicted scaling limit in \eqref{susy_singularity1}.
In Fig.\ \ref{globalsol2dsusy} we show the plot of the on-shell effective potential extending beyond $\tilde{\varphi}$ for the LPA' supersymmetric Ising class 
of Subsect.\ \ref{subsection_susy_ising2d}. 
The extended solution does not encounter further singularities beyond $\tilde{\varphi}$ even though its asymptotic behavior is governed by \eqref{susy_singularity1},
which is very similar to the asymptotic behavior
of scalar field theory in \eqref{singularity0}.

The natural question to ask at this point is th following: Does the extension of the critical solution affect the spectrum of its fluctuations?
It is possible to show, in general, that this is not the case. Let us consider the linearized equation for the fluctuations $\delta w'(\varphi)$ of $w'(\varphi)$ around
the critical solution, which is obtained by deforming
\begin{equation}
 w'(\varphi) \to w^{\prime *}(\varphi) + \epsilon\, \delta w'(\varphi) \left(\frac{k}{k_0}\right)^{-\theta}
\end{equation}
in the renormalization group flow $k\partial_k w'(\varphi)$ which can be easily obtained by applying a $\varphi$ derivative to both sides of \eqref{susy_wdot}.
The linearized ODE is of the form
\begin{equation} \label{wprime_fluctuations}
 \left(f(\varphi)\partial^2_\varphi +g(\varphi) \partial_\varphi +\frac{d-\eta}{2}+\theta\right)\delta w'(\varphi)=0\,,
\end{equation}
in which we introduced the functions
\begin{equation}
\begin{split}
  f(\varphi)&={\cal F}'_{w,\eta}(w'')\,,\\
  g(\varphi)&={\cal F}''_{w,\eta}(w'')w'''(\varphi)+\frac{d-2+\eta}{2}\varphi\,,
\end{split}
\end{equation}
which have to be evaluated at the critical solution $w^{\prime *}(\varphi)$.
As we did in \eqref{transformed_fluctuation} for the simple scalar model,
we transform the fluctuations in \eqref{wprime_fluctuations}
so that the transformed equation does not contain first derivatives with respect to the field.
This is done by defining the new fluctuations $\delta u(\varphi)$ such that
\begin{equation}\label{transformation12}
\delta u(\varphi) = \exp\Bigl(\frac{1}{2}\int_{\varphi_0}^\varphi {\rm d}y\, \frac{g(y)}{f(y)}\Bigr)\delta w'(\varphi)\,,
\end{equation}
where we can choose $\varphi_0>\tilde{\varphi}$ or $\varphi_0<\tilde{\varphi}$
for the lower integration limit.
In both cases, the transformed fluctuations solve a 
Schr\"odinger-type equation of the form
\begin{equation}\label{transformed12}
 \left(f(\varphi)\partial_\varphi^2 +P(\varphi)+\theta\right)\delta u(\varphi)=0\,.
\end{equation}
Since $f(\tilde\varphi)=0$ and 
$g(\tilde \varphi)=(1-\eta)w''(\tilde\varphi)/w'''(\tilde\varphi)>0$
the transformation \eqref{transformation12} is singular at the point
$\tilde\varphi$.
Since in addition $f(\varphi)$ has positive slope at $\tilde\varphi$
it follows that the exponential factor in \eqref{transformation12}
vanishes when $\varphi$ approaches $\tilde\varphi$ from below and from above.
This means that $\delta u$ obeys Dirichlet boundary conditions at
$\varphi=\tilde\varphi$.
It follows that the spectral problem \eqref{transformed12} 
has two disjoint sets of solutions, solutions with $\varphi<\tilde{\varphi}$ and solutions with $\varphi>\tilde{\varphi}$.
The only admitted fluctuations satisfy either the boundary 
conditions \eqref{wp_bc_odd_w} or \eqref{wp_bc_even_w},
which are possible only for the transformed fluctuations for which $\varphi_0<\tilde{\varphi}$, thus implying that $\varphi_0=0$ is the most natural choice.
We explored numerically the spectrum of the global solution plotted
in Fig.\ \ref{globalsol2dsusy} using the SLAC derivative method discussed in
Appendix \ref{appendix_slac} applied to the eigenvalue problem \eqref{transformed12}.
It can be checked that the spectrum does not receive any correction from 
the extension of the 
domain of existence of the fixed-point solution beyond $\tilde\varphi$.

\end{appendix}

\vfill

\end{document}